\documentclass[twocolumn]{aastex701}

\newcommand\fclumpy{$f_{\rm{clumpy}}$}

\usepackage{rotating}

\begin{document}

\title{The Fraction of Clumpy Galaxies in JADES Over $2<z<9$}

\correspondingauthor{Alexander de la Vega}

\author[0000-0002-6219-5558, gname=Alexander, sname='de la Vega']{Alexander de la Vega}
\affiliation{Department of Physics and Astronomy, University of California, 900 University Avenue, Riverside, CA 92521, USA}
\email{alexandd@ucr.edu}

\author[0000-0001-5846-4404]{Bahram Mobasher}
\affiliation{Department of Physics and Astronomy, University of California, 900 University Avenue, Riverside, CA 92521, USA}
\email{mobasher@ucr.edu}

\author[0000-0002-0364-1159]{Zahra Sattari}
\affiliation{IPAC, California Institute of Technology, MC 314-6, 1200 E. California Boulevard, Pasadena, CA 91125, USA}
\email{zsattari@ipac.caltech.edu}

\author[0000-0003-3691-937X]{Nima Chartab}
\affiliation{IPAC, California Institute of Technology, MC 314-6, 1200 E. California Boulevard, Pasadena, CA 91125, USA}
\email{nchartab@ipac.caltech.edu}

\author[0009-0008-4976-3216]{Faezeh Manesh}
\affiliation{Department of Physics and Astronomy, University of California, 900 University Avenue, Riverside, CA 92521, USA}
\email{fakhl001@ucr.edu}

\author[0009-0009-9776-6849]{Niloofar Sharei}
\affiliation{Department of Physics and Astronomy, University of California, 900 University Avenue, Riverside, CA 92521, USA}
\email{niloofar.sharei@email.ucr.edu}



\begin{abstract}

\noindent High-redshift galaxies exhibit compact regions of intense star formation, known as ``clumps,'' which are conspicuous in the rest-frame ultraviolet. Studying them can shed light on how they form and evolve and inform theoretical models of galaxy evolution. We examine the evolution of clumpy galaxies with redshift and stellar mass over ${2<z<9}$ with James Webb Space Telescope (JWST) imaging from the JWST Advanced Extragalactic Survey. Off-center clumps are detected in the rest-frame near-ultraviolet using similar techniques to those in earlier studies based on Hubble Space Telescope (HST) images. This is done for a sample of 9,121 star-forming galaxies with stellar masses $\log\left(M_{\star}/M_{\odot}\right) \geq 8$. The fraction of clumpy galaxies, \fclumpy, increases from ${\sim10\%}$ at ${z\sim7.75}$ to ${\sim70\%}$ at $z\sim2.75$ at $\log\left(M_{\star}/M_{\odot}\right) \geq 9$. Our \fclumpy~values are generally higher at fixed redshift and increase faster with decreasing redshift than what studies based on HST data found, which we attribute largely to the higher sensitivity of JWST. \fclumpy~correlates with stellar mass. Our \fclumpy~measurements are compared with those from simulations as well as other observations. At low redshifts ($z\lesssim5.75$) and intermediate-to-high stellar masses ($\log\left(M_{\star}/M_{\odot}\right) \geq 9$), our results suggest gas fragmentation due to violent disk instabilities to be the dominant mechanism for forming clumps. At high redshifts and {intermediate} stellar masses, compression of gas during mergers appears to dominate. 

\end{abstract}



\section{Introduction}

Early Hubble Space Telescope (HST) imaging of high-redshift galaxies revealed systems with patchy, clumpy, kiloparsec-scale substructures \citep{Cowie95, Giavalisco96, vdB96, Elmegreen05, Elmegreen07}. Later work showed that these substructures (hereafter, ``clumps'') are bright in the rest-frame ultraviolet (UV), are fairly massive ($\gtrsim10^7~M_{\odot}$), have young stellar ages ($\sim100$ Myr), and, {on average,} make up about $10-50\%$ of the total star formation rates (SFRs) and about $4-20\%$ of the total stellar mass of their host galaxies \citep{Elmegreen08, Elmegreen09, ForsterSchreiber11, Guo12, Guo15, Guo18, Wisnioski12, Wuyts12, Wuyts13, Livermore15, Shibuya16, Soto17, Zanella19, Mehta21, Vanzella21, Mestric22, Claeyssens25, Sok25}. These results suggest that clumps were significant sites of star formation in galaxies over $0.5<z<3$, during the peak of cosmic star formation activity \citep{MadauDickinson14}. 

Studies of observed and simulated clumps have proposed two main formation mechanisms. The first is violent disk instabilities (VDI), wherein gas collapses under gravitational instabilities in marginally stable gaseous disks \citep{Noguchi98, Noguchi99, Immeli04_ism, Immeli04_hst, Bournaud07, Bournaud09, Agertz09, Dekel09_theory, Ceverino10, Romeo10, Romeo14, Inoue16}. The VDI is sustained by the high gas fractions in high-redshift galaxies \citep[][and references therein]{Tacconi20}, fed by the accretion of cold gas from the intergalactic medium \citep{Birnboim03, Keres05, Keres09, Dekel06, Bournaud09_accretion, Dekel09, Dekel09_theory, Nelson13}. Observational support for the VDI {comes from} integral field spectroscopic data, with which maps of gas kinematics may be measured \citep{ForsterSchreiber11, Genzel11, Genzel14, Wisnioski11, Wisnioski12, Newman12, Mieda16}. The second mechanism is mergers, during which gas is compressed by tidal forces and/or turbulence due to gravitational interactions between galaxies \citep{DiMatteo08, Renaud15, Nakazato24}. Observational support for mergers has come from visual inspection of the morphologies of clumpy and nonclumpy galaxies \citep{Straughn15} and integral field spectroscopy \citep{Puech10, MenendezDelmestre13}.

Clumps could also evolve within their host galaxies and alter their morphologies. Theoretical studies and simulations have predicted that clumps experience dynamical friction and/or interactions, which cause them to migrate to the centers of their hosts and contribute to the growth of bulges \citep{Noguchi99, Immeli04_ism, Immeli04_hst, Dekel09_theory, Dekel22, Krumholz10, Ceverino12, Inoue12, Perez13, Bournaud14, Mandelker14, Mandelker17}. However, other simulations find that clumps are short-lived and dissipate within $\sim100$ Myr due to stellar feedback, which drives strong outflows of gas (\citealt{Murray10, Genel12, Hopkins12, Moody14, Buck17, Oklopcic17, Meng20}; see also \citealt{Fensch21}). {
 Regardless whether clumps migrate to the centers of their hosts,} stars formed within the clumps could {also} contribute to the growth of the thick disk (\citealt{Noguchi96, Bournaud07, Inoue14}).

The evolution of the fraction of clumpy galaxies (\fclumpy) with redshift may also provide clues to the physical processes through which clumps form and evolve \citep{Guo15, Shibuya16}. Using HST imaging data, \fclumpy~has been estimated at various redshifts by identifying compact, UV-bright regions within galaxies. The behavior of \fclumpy~with redshift is well understood at $z\leq3$, but unclear at higher redshifts. At $z\leq3$, \fclumpy~peaks at $\sim60\%$ at $z\sim2$ and declines at lower and higher redshifts \citep{Elmegreen07, Overzier09, Puech10, Guo12, Murata14, Guo15, HinjosaGoni16, Shibuya16, Adams22, Sok22, Martin23, Sattari23}. These studies are largely consistent with predictions from simulations of the  redshift evolution of galaxies undergoing VDI \citep{Cacciato12}. Support for mergers was also found at lower redshift \citep{Puech10} and lower masses \citep{Guo15}. At $z>3$, there is disagreement on the behavior of \fclumpy~with redshift{, with one set of observations finding that \fclumpy~decreases from $\sim40\%$ at $z\sim3$ to $\sim20\%$ at $z\sim8$ \citep{Ravindranath06, Conselice09, Oesch10, Shibuya16}, and another finding \fclumpy~to be roughly constant with redshift at $\sim40\%$ from $z\sim4$ to $z\sim8$ \citep{Jiang13,  Kawamata15, Curtis-Lake16, Bowler17, Ribeiro17}}.

Studies using data from the James Webb Space Telescope (JWST) have revealed a more complex picture. Thanks to the higher angular resolution, greater sensitivity, and longer wavelength coverage of the JWST, subkiloparsec substructures in faint galaxies at high redshift can now be studied in greater detail with its Near-Infrared Camera (NIRCam). JWST imaging of gravitationally lensed galaxies over $1 < z < 8.5$ revealed a significant population of clumps that were not detected by the HST \citep{Claeyssens23, Fujimoto25}. Unlensed galaxies at $z>4$ also reveal clumps in JWST images \citep[e.g.,][]{Tacchella23, Hainline24}, some of which are not found in HST observations \citep[e.g.,][]{Bik24, Tanaka24}. On average, 76\% of the rest-frame UV light from $z\simeq6-8$ galaxies was found to come from the clumps \citep{Chen23}, which is much higher than the $\sim20-30\%$ measured over $0.5<z<3$ \citep{Guo15}. Finally, about 70\% of spectroscopically selected UV-bright galaxies at $z\sim7$ contain clumps \citep{Harikane25}, in contrast with the $10-20$\% inferred from pre-JWST studies \citep[][and references therein]{Shibuya16}. 

While recent studies show that clumpy galaxies might be more abundant than expected at high redshifts, interpretation of these results is difficult due to selection effects and occasional small sample sizes. The largest post-JWST sample at $z>6$ to date for which \fclumpy~has been measured consists of 53 galaxies \citep{Harikane25}. That sample was selected to be intrinsically bright with absolute UV magnitudes $-23 < M_{UV} < -19$ mag. \citet{Claeyssens25} examined the physical properties of 1,751 clumps in 377 galaxies over $0.7 < z < 5.5$ that are gravitationally lensed by the cluster A2744. A measurement of \fclumpy~as a function of redshift was not performed in that work, {due to the intrinsic incompleteness of studies on lensed galaxies. \citet{Tan25} measured \fclumpy~over $0.25<z<5$ for a sample of 877 galaxies expected to be progenitors of Milky Way analogs. Their sample was selected with bins in stellar mass that evolve with redshift. This complicates comparisons with pre-JWST studies, which largely selected galaxies with a mass range that was constant with redshift. \citet{Sok25} measured \fclumpy~ and stellar masses of clumps in $\approx4,500$ lensed galaxies with magnifications $\mu \leq 2$.}

In this paper, we measure \fclumpy~over ${2<z<9}$ with JWST data for a sample of 9,121 star-forming galaxies. The selection of the sample is performed such that results from the JWST and HST data may be directly compared. This allows us to determine whether the redshift evolution of \fclumpy~is significantly different {between HST and JWST data}. 

The paper is organized as follows. In section \ref{sec:data}, we describe the data used in this work. The selection criteria of our sample are given in section \ref{sec:sample_selection}. section \ref{sec:selection_of_clumpy_galaxies} {describes the routine used to identify clumpy galaxies and its completeness as a function of redshift and stellar mass}. { In section \ref{sec:nature_clumps}, the physical nature of the identified clumps is discussed.} The redshift evolution of \fclumpy~is discussed in section \ref{sec:clumpy_fraction}. In section \ref{sec:clumpy_fraction_mass}, we examine the correlation between stellar mass with \fclumpy. Predictions from simulations of clumpy galaxies are compared with our observations in section \ref{sec:discuss_sims}. We discuss the implications for clump formation mechanisms in section \ref{sec:discuss_obs}. We summarize and conclude in section \ref{sec:conclusions}. 

Throughout this work, cosmological parameters measured by \citet{Planck16} are adopted, with $H_0 = 67.7~\rm{km}~\rm{s}^{-1}~\rm{kpc}^{-1}, \Omega_{\rm{m}} = 0.309$, and $\Omega_{\Lambda} = 0.691$. All magnitudes are in the AB photometric system \citep{Oke83}. 

\section{Data}
\label{sec:data}

\subsection{Images, Source Detection, and Photometry}
\label{sec:photometry}
This work is based on NIRCam observations of the Great Observatories Origins Deep Survey (GOODS; \citealt{Giavalisco04}) North and South (hereafter, GOODS-N and GOODS-S) fields, taken {as part of Data Release 3 of} the JWST Advanced Deep Extragalactic Survey (JADES; \citealt{Rieke23, Eisenstein26, Eisenstein25, Bunker24, DEugenio25}). These fields were also observed by HST as part of the Cosmic Assembly Near-infrared Deep Extragalactic Legacy Survey \citep[CANDELS;][]{Grogin11, Koekemoer11}. The combined area of these fields observed by NIRCam is 124 arcmin$^2$. The GOODS-N field was observed in 11 NIRCam bandpasses: F090W, F115W, F150W, F182M, F200W, F210M, F277W, F335M, F356W, F410M, and F444W. GOODS-S was observed in these and three additional bandpasses: F430M, F460M, and F480M. Observations in GOODS-S in the F182M, F210M, F430M, F460M, and F480M bandpasses were taken by the JWST Extragalactic Medium-Band Survey \citep[JEMS;][]{Williams23}. The 5$\sigma$ point-source limiting magnitude measured in 0.\arcsec3 circular apertures is $\sim29.5$ mag in each bandpass. 

{Source detection and photometry in nine NIRCam and 14 ground- and space-based bandpasses in JADES was described in \citet{Rieke23}}. Briefly, a detection image was created by {convolving the F277W, F335M, F356W, F410M, and F444W images to the resolution of the F444W image, stacking them,} and weighting them by their inverse variances. Next, various source detection routines in the {\tt python} module {\tt Photutils} \citep{Bradley23} were used to detect and deblend sources and perform photometry. All images are aligned to the same reference frame and drizzled to a common pixel scale of 0.\arcsec03 \ per pixel. {Photometry was then performed} on HST and JWST images convolved to the F444W resolution. For more details, see section 4 in \citet{Rieke23}. 

This work relies on photometry in HST and JWST bandpasses. Spatially integrated fluxes are measured in Kron apertures with a Kron parameter of $k=2.5$. Spatially integrated errors on the HST and JWST photometry are computed as follows. Two sources of error are considered: errors due to sky subtraction and a combination of source Poisson and instrumental errors. The former are measured in apertures of various sizes that are centered on empty regions of the sky (see section 4.2 in \citealt{Rieke23}, also \citealt{Labbe05}; \citealt{Whitaker11}; and \citealt{Skelton14}). For JWST/NIRCam images, the latter come from the ERR extension, and for HST images, they are measured following the procedure described in Sec. 4.1 by \citet{vdW12}. Total errors on the JWST photometry are computed by adding these two sources of error in quadrature. For the HST images, the JADES team have measured errors due to sky subtraction and report the average values of the weight, or inverse variance, within the Kron apertures. The error due to the weight for each galaxy is computed by multiplying the average weight value by the area of the Kron aperture in pixels. Total errors on the HST photometry are calculated by adding these two errors in quadrature. A final error is added to all fluxes for each bandpass to account for zero-point errors and mismatches in the templates between model and observed SEDs. It is equal to 5\% of the flux (see, e.g., Table 4 of \citealt{Dahlen13} and Table 11 of \citealt{Skelton14}). 

Error maps in the JWST bandpasses, which are needed to detect clumps, come from the ERR extension in the images provided by the JADES team. Those in the HST bandpasses are measured following the procedure described in Sec. 4.1 in \citet{vdW12}. 

Axis ratios, half-light radii, segmentation maps, and central coordinates of galaxies were measured from the detection images by \citet{Rieke23}. These measurements and data products are adopted for this analysis. They were calculated by applying {\tt photutils} to the detection image in each field. Half-light radii were measured in circular apertures. Central coordinates were computed by taking the light-weighted centroids along the horizontal and vertical axes of the detection images. 

\subsection{Photometric and Spectroscopic Redshifts}
\label{sec:redshifts}
Photometric redshifts in JADES come from \citet{Hainline24} and are estimated using {\tt EAZY} \citep{Brammer08}. {\tt EAZY} measures photometric redshifts by fitting nonnegative linear combinations of templates to observed fluxes in order to derive probability distribution functions of the redshift. Templates are supplied by the user. \citet{Hainline24} develop modified versions of the original galaxy templates used by \citet{Brammer08} and seven new templates designed to fit the spectral energy distributions (SEDs) of high-redshift galaxies. For more details, see section 3 by \citet{Hainline24}. 

Spectroscopic redshifts come from JADES and the compilation over the CANDELS fields by \citet{Kodra23}. Descriptions of the selection of the targets, reduction and calibration of the data, measurements of the redshifts, and presentation of the JADES spectra are provided by \citet{Bunker24} and \citet{DEugenio25}. The provenances of the spectroscopic redshifts gathered by \citet{Kodra23} and descriptions of the quality cuts applied can be found in that work and references therein.

\begin{figure}
    \centering
    \includegraphics[width=\linewidth]{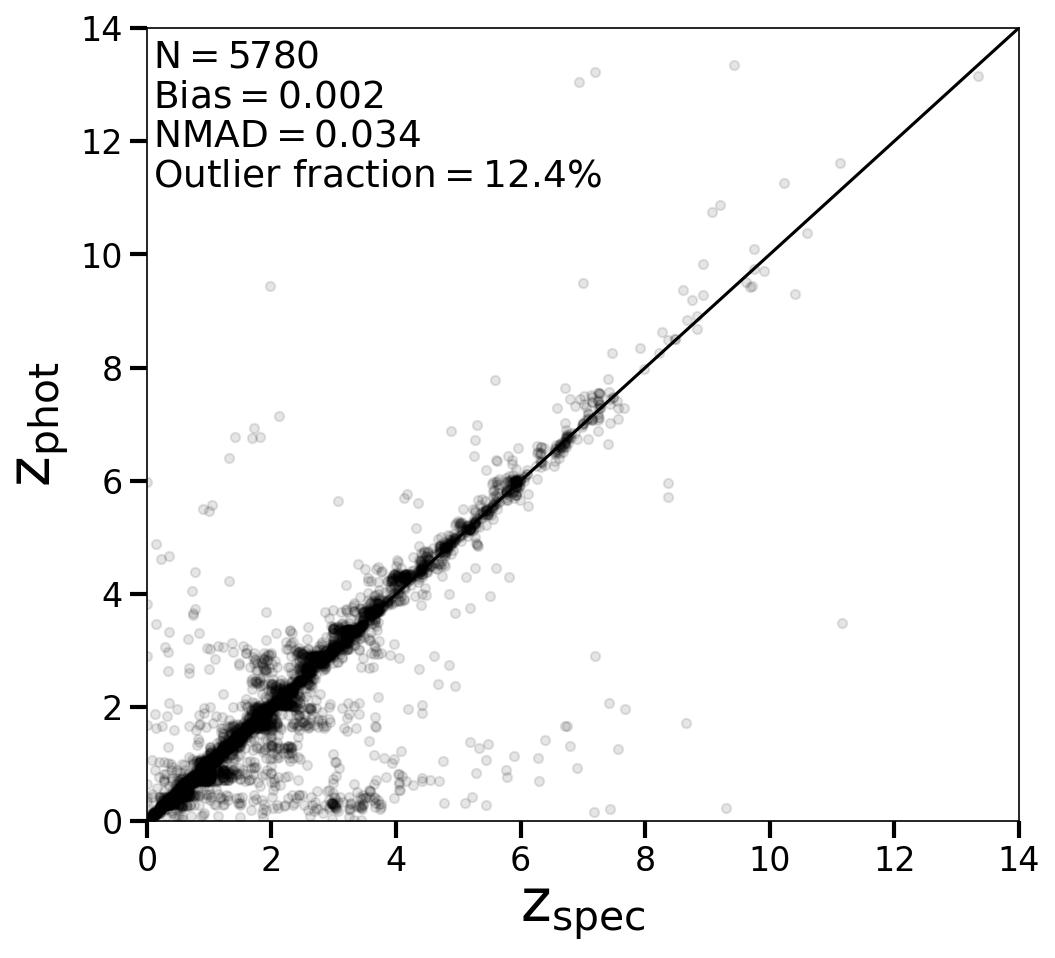}
    \caption{Comparison of 5,780 spectroscopic and photometric redshifts in the JADES fields. The former are compiled from various sources (see section \ref{sec:redshifts}) and the latter are measured using {\tt EAZY} \citep{Brammer08}. The bias, normalized median absolute deviation (NMAD) scatter, and outlier fraction are measured according to \citet{Dahlen13}. }
    \label{fig:phot_spec_z_comp}
\end{figure}

The accuracy of the photometric redshifts measured by the JADES team is assessed by comparing them with spectroscopic redshifts. We find 2,774 {(3006) with} spectroscopic redshifts in GOODS-N {(GOODS-S)}. Their redshifts span $0<z<13.35$. The accuracies of the photometric redshifts are quantified using the bias, normalized median absolute deviation (NMAD) scatter, and outlier fraction according to \citet{Dahlen13}. The bias is given by 
\begin{equation}
    \textrm{bias} = \textrm{mean} \left(\Delta z / (1+z_{\rm{spec}})\right),
\end{equation}
where $\Delta z = z_{\rm{spec}} -z_{\rm{phot}}$. It is computed after excluding outliers, which are defined below. The NMAD scatter is:
\begin{equation}
    \textrm{NMAD} = 1.48 \times \textrm{median} \left( |\Delta z| / (1+z_{\rm{spec}}) \right).
\end{equation}
The outlier fraction is given by the fraction of galaxies that satisfy $|\Delta z | / \left(1+z_{\rm{spec}})\right) > 0.15$. 

Spectroscopic and photometric redshifts are compared in Figure \ref{fig:phot_spec_z_comp}. We measure a bias of 0.002, an NMAD scatter of 0.034, and an outlier fraction of 12.4\% for all 5,780 redshifts from both fields. There is good agreement between the spectroscopic and photometric redshifts, given the low bias and NMAD scatter. The outlier fraction is somewhat high, which may be due to the shallower depth of and fewer medium bands available in GOODS-N than those of GOODS-S. Similar NMAD scatter and outlier fractions were obtained by \citet{Puskas25}. They built a catalog of spectroscopic redshifts in GOODS-N and GOODS-S and classified them in terms of their quality. Of these, 5,382 in GOODS-S were found to have the highest quality, for which they measured an NMAD scatter of 0.065 and outlier fraction of 6.91\%. For the 2,591 highest-quality redshifts in GOODS-N, an outlier fraction of 13.62\% was measured. 

\begin{figure*}[t!]
    \includegraphics[width=\textwidth]{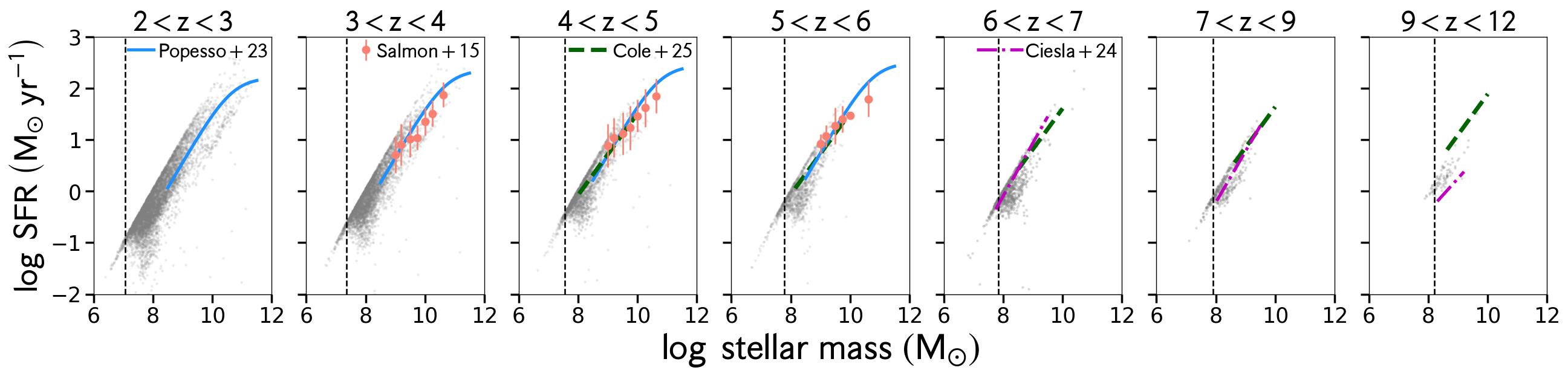}
    \caption{The relation between stellar mass and SFR, also known as the star formation main sequence (SFMS), for the sample {of galaxies} (section \ref{sec:sample_selection}) over $2<z<12$. These quantities are measured from fits to the combined HST and JWST SEDs using {\sc bagpipes} \citep[][see section \ref{sec:masses}]{Carnall18}. Measurements for individual galaxies are shown as gray dots in various redshift bins, a different one of which is in each {panel}. Completeness limits in stellar mass {in each redshift bin} are indicated with vertical dashed lines. There is broad agreement between the SFMS presented here and that measured in other studies. The SFMS measured by \citet{Popesso23} over $2<z<6$ from a compilation of studies is shown as a blue solid line in the leftmost four {panels}. That measured by \citet{Salmon15} over $3.5 < z < 6.5$ is plotted in the second through fourth columns. The SFMS from \citet{Cole25}, measured over $4.5 < z < 12$, is shown as a green dashed line. The SFMS measurements by \citet{Ciesla24} over $6<z<12$ are shown as magenta dashed-dotted lines.}
    \label{fig:sfms}
\end{figure*}

\subsection{Stellar Masses and Star Formation Rates}
\label{sec:masses}

Stellar masses and star formation rates for all galaxies in the full sample (section \ref{sec:sample_selection}) are measured by fitting their SEDs. The SFRs used in this work are {taken to be the average values inferred from the most recent 100 Myr in the fit star formation histories (SFHs)}. Where available, fits were performed to the HST/Advanced Camera for Surveys (ACS) bandpasses F435W, F606W, and F775W combined with all the NIRCam bandpasses listed in section \ref{sec:photometry}. The number of filters used in each fit ranges from 5 to 17, with a median of 13. The Bayesian SED-fitting tool {\sc bagpipes} \citep{Carnall18} is used. It fits SEDs using the {\sc MultiNest} Markov Chain Monte Carlo algorithm \citep{Feroz09} and returns marginalized posterior probability distribution functions for each parameter of interest. Stellar masses and SFRs are measured by taking the median stellar mass and SFR from their respective posteriors.

We adopt the priors listed in Table 1 in \citet{Carnall2020}, adjusting those on the metallicity and the amount of dust attenuation in the rest-frame $V$-band, $A_V$. This is done based on arguments for physically motivated priors given by \citet{delavega25}. A Gaussian prior on the metallicity $Z/Z_{\odot}$ with a mean $\mu=0.3$ and standard deviation $\sigma=0.5$ is assumed. Another Gaussian prior on $A_V$ is assumed, with $\mu=0.3$ and $\sigma=1.0$, over the range $A_V \in [0, 6]$. This is adopted from \citet{Leja19}. 

Priors for other physical parameters are as follows. A uniform prior on the log of the stellar mass $\log\left(M_{\star}/M_{\odot}\right)$ is assumed over the range of 1 to 13. We adopt the flexible attenuation curve model by \citet{Salim18}. It modifies the shape of the attenuation curve measured by \citet{Calzetti00} by multiplying it by a power law that varies with wavelength. For the slope of the power law, $\delta$, a Gaussian prior is set with $\mu=0$ and $\sigma=0.1$ over the range [-0.3, 0.3]. The \citet{Calzetti00} curve is obtained by setting $\delta = 0$. The strength of the dust absorption feature at rest-frame 2,175 \AA \ is also modeled by \citet{Salim18}. We set a uniform prior on it over the range [0, 5]. Another prior is assumed for the differential attenuation between young ($\leq10$ Myr) and old stars \citep[see, e.g.,][]{Calzetti00, CF00}. The former are generally dustier than the latter due to additional attenuation from birth clouds. A uniform prior on $\eta$, the ratio of $A_V$ for young stars to that for old stars, is assumed over the range [1, 5]. SED fits include nebular emission, the strength of which is determined with the log of the ionization parameter, $\log U$, and the nebular metallicity. A uniform prior is set on the former over the range [$-4, -2$]. The latter is assumed to be the same as that of the stars. 

Last, a double power-law {SFH} is assumed{,} given by 
\begin{equation}
    \textrm{SFR}(t) \propto \left[ \left(\frac{t}{\tau}\right)^{\alpha} + \left(\frac{t}{\tau}\right)^{-\beta} \right]^{-1},
\end{equation}
where $t$ is the time since star formation began in a galaxy, $\tau$ is the time after the Big Bang when the SFR peaked, $\alpha$ is the declining slope, and $\beta$ is the rising slope. We assume log-uniform priors for $\alpha$ and $\beta$, each of which is over the range [0.01, 1000]. A uniform prior on $\tau$ is set, which ranges from 0.1 Gyr to the age of the Universe at the observed redshift of a given galaxy. 

For the SED fits, we fix the redshift to its spectroscopic value, if available. Otherwise, the photometric redshift from JADES is assumed. We apply upper limits on photometry with signal-to-noise ratio (S/N) $< 3$ following the procedure by \citet{delavega25}. This is relevant only for the HST and bluest NIRCam filters, especially at high redshift. 

Stellar population synthesis templates from version 2.2.1 of the Binary Population and Spectral Synthesis \citep[BPASS;][]{Eldridge17, StanwayEldridge18} models are used. The initial mass function is set to the default `135\_300' model listed in Table 2 in \citet{StanwayEldridge18}. The {\sc cloudy} \citep{Ferland17} photoionization code is used to model nebular emission. All HST and JWST filter transmission curves, which are used in the SED fits, are taken from the SVO Filter Profile Service \citep{Rodrigo12, Rodrigo20}. 

In Figure \ref{fig:sfms}, the relation between stellar mass and SFR, also known as the star formation main sequence \citep[SFMS; e.g.,][]{Noekse07, Rodighiero10, Whitaker14, Salmon15, Schreiber15, Tomczak16, Popesso23, Ciesla24, Cole25} is shown. These properties are derived from the SED fits. The distribution of stellar masses and SFRs measured in this work agrees broadly with the SFMS measured in various studies. 

\subsection{Stellar Mass Completeness Limits}
\label{sec:mass_completeness}

Here we compute 90\% mass completeness limit for our selected sample. The limits are based on the method in \citet{Pozzetti10} and follow what was done for galaxies in JADES by \citet{Simmonds24}. The stellar masses obtained with {\sc bagpipes} are used. 

The completeness limits are estimated as follows. First, a limiting stellar mass, $M_{\rm{lim}}$, is computed for all galaxies in the full sample (section \ref{sec:sample_selection}). It is defined to be
\begin{equation}
    \log \left(M_{\rm{lim}}\right) = \log\left(M_{\star}\right) + 0.4\left(m-m_{\rm{lim}}\right),
\end{equation}
where $\log\left(M_{\star}\right)$ is the stellar mass in solar units of a given galaxy, $m$ is its apparent magnitude, and $m_{\rm{lim}}$ is the limiting magnitude of the survey. In this work, the magnitude in the F444W band is used to calculate $M_{\rm{lim}}$. This band is used as it is the longest{-wavelength} NIRCam filter available in both GOODS-N and GOODS-S and thus best traces the stellar mass. After finding $M_{\rm{lim}}$ for all galaxies, {we divide them into redshift bins} (section \ref{sec:sample_selection}). In each bin, the faintest 20\% of galaxies are identified. Then, the 90th percentile of $M_{\rm{lim}}$ for these is found. We take $m_{\rm{lim}}$ to be 29.8, which is the average $5\sigma$ point-source depth across the GOODS-N and GOODS-S fields, as measured by \citet{Puskas25}.

Completeness limits span $\log \left(M_{\star}/M_{\odot}\right) = 7.16 - 8.21$. They change monotonically with redshift. These limits are comparable to those derived by other studies where there is overlap in redshift \citep[e.g.,][]{Ciesla24, Simmonds24, Puskas25}.

\begin{figure}[t!]
    \centering
    \includegraphics[width=\linewidth]{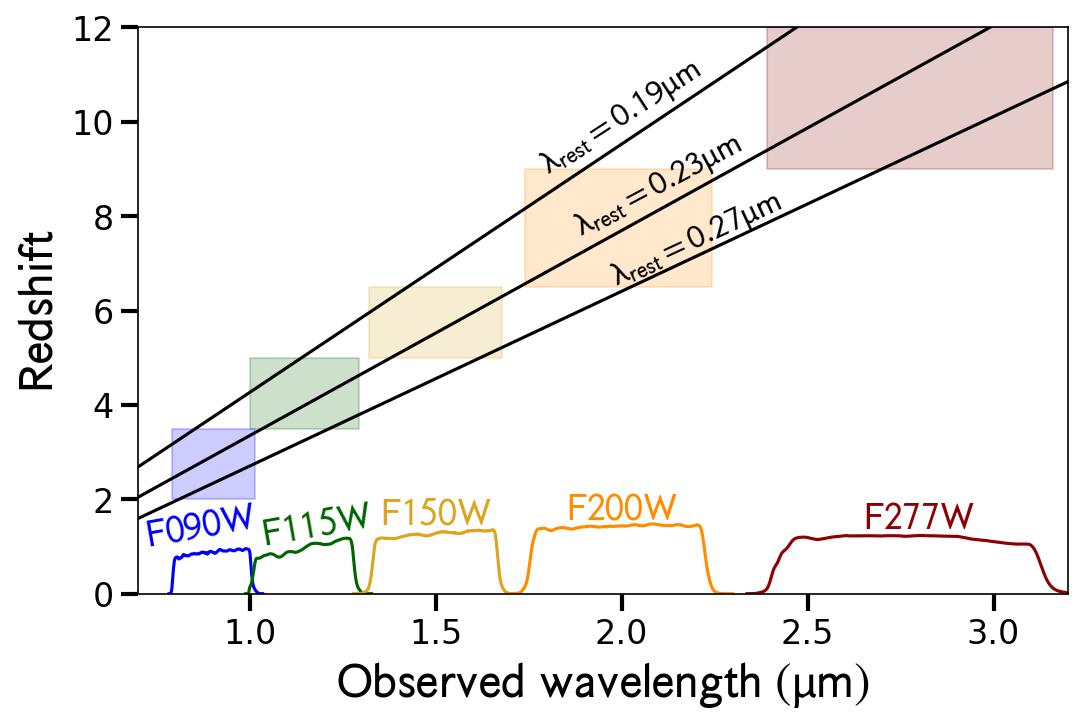}
    \caption{Clumps over redshifts $2<z<12$ are identified in the rest-frame near-UV, following pre-JWST studies. Bandpasses are chosen such that the rest-frame $0.19 - 0.27~\mu$m is sampled. Shown here are the bandpasses used to find clumps at various redshifts. The selected filters used to find the clumps are F090W over $2<z<3.5$; F115W over $3.5<z<5$; F150W over $5<z<6.5$; F200W over $6.5 <z <9$; and F277W over $9<z<12$. These ranges in redshift are represented by the heights of the shaded regions. Their widths correspond to wavelength ranges over which the transmission curves of the bandpasses, shown at the bottom, are at least 5\% of the maximum transmission. Shaded regions and transmission curves share the same colors. }
    \label{fig:filter_rf_nuv}
\end{figure}

\section{Selection of the Sample of Galaxies}
\label{sec:sample_selection}

We start with all sources in the JADES GOODS-N and GOODS-S photometric catalogs. There are 179,709 sources in both fields: 85,709 in GOODS-N and 94,000 in GOODS-S. 

All sources that are brighter than or equal to 29th magnitude in the reddest broadband filter, F444W, are selected. This is about 0.5 mag brighter than the limiting magnitudes for all NIRCam filters in JADES \citep[Table 2 in][]{Rieke23}. This cut removes 82,606 objects with 97,103 remaining. Next, we remove point sources using the {\tt FLAG\_ST} flag. This cut removes 531 sources, leaving 96,572 galaxies. All sources that are contaminated by diffraction spikes from bright stars are removed with the {\tt FLAG\_BS} flag. Those that are contaminated by bright neighbors are removed using the {\tt FLAG\_BN} flag. The former cut removes 1,504 galaxies and the latter removes 23,672, with 71,396 left. 

Galaxies over $2<z<12$ are selected. This is done to enable the identification of clumps in the rest-near-ultraviolet (NUV) at all redshifts, following pre-JWST studies \citep[e.g.,][]{Wuyts12, Guo15, Shibuya16}, over the widest possible range in redshift. The lower limit of $z=2$ is chosen because the bluest NIRCam bandpass available, F090W, spans the rest-frame NUV starting at $z=2$. The upper limit of $z=12$ is chosen such that the reddest bandpasses sample the rest-frame Balmer break. The range $2<z<12$ thus spans the rest-UV to rest-optical at all redshifts for the NIRCam bandpasses in JADES. This rest-frame wavelength coverage is optimal for the measurement of stellar masses and SFRs \citep{Pacifici12, Pforr12, Buat14}. Applying the redshift cut removes 38,947 galaxies and leaves 32,449. 

Further cuts are applied to select galaxies that are brighter than or equal to 29th magnitude in a NIRCam bandpass spanning the rest-frame NUV, which is assumed to cover 0.19 -- 0.27 $\mu$m (see the GALEX NUV bandpass). Galaxies must be brighter than 29th magnitude in F090W over $2<z<3.5$; in F115W over $3.5<z<5$; in F150W over $5<z<6.5$; in F200W over $6.5 <z <9$; and in F277W over $9<z<12$. These redshift ranges and choices of bandpasses are shown in Figure \ref{fig:filter_rf_nuv}. Applying these cuts removes 13,256 galaxies and leaves 19,193. 

Cuts are then applied to exclude compact and edge-on galaxies{, following} pre-JWST studies of clumpy galaxies \citep[e.g.,][]{Guo12, Guo15, Shibuya16, Sattari23}. Galaxies with circular half-light radii of at least 0.\arcsec1 in the F444W bandpass are selected. {This minimum size is slightly larger than the half width at half maximum of the F444W point-spread function (PSF, \citealt{ZhuangShen24}) and is imposed in order to select galaxies that are resolved.} This removes 2,848 galaxies with 16,345 remaining. Galaxies with axis ratios of $b/a \geq 0.5$ are chosen. This excludes 933 more {galaxies}. 

\begin{figure}[t!]
    \centering
    \includegraphics[width=\linewidth]{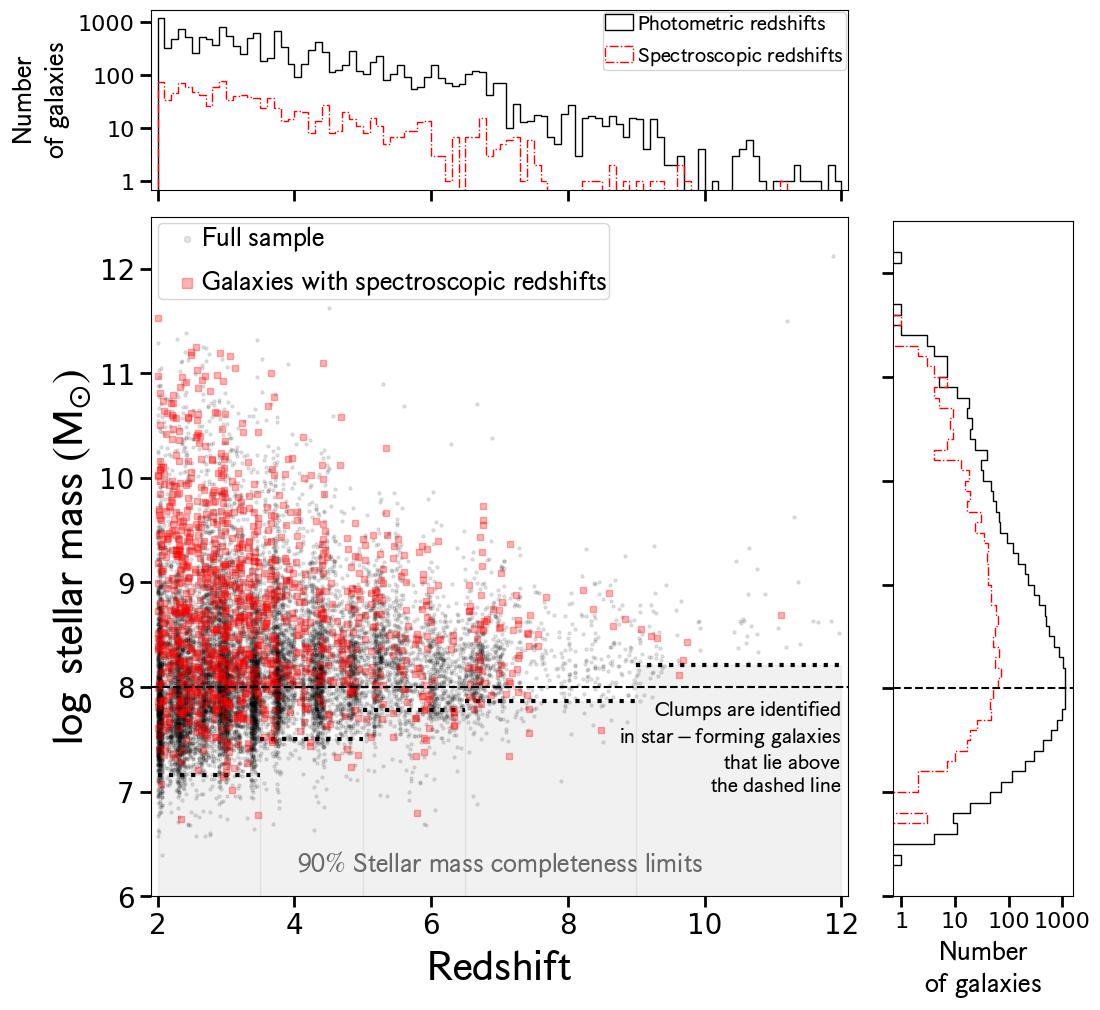}
    \caption{The stellar masses as a function of redshift are shown for the full sample (section \ref{sec:sample_selection}). Individual galaxies with only photometric redshifts are plotted as small gray dots and those with spectroscopic redshifts are indicated with red squares. Top panel: the distribution of redshifts. Right panel: stellar mass distribution. In the top and right panels, the distributions for galaxies with spectroscopic redshifts are denoted with red dashed-dotted histograms. Clumps are identified in galaxies with stellar masses of at least $10^8~M_{\odot}$, which is indicated with the horizontal dashed line. It is higher than or close to the 90\% stellar mass completeness limit of the sample at all redshifts, which are shown as light gray shaded regions and capped with dotted lines.}
    \label{fig:sample_completeness}
\end{figure}

The full sample contains 15,412 galaxies: 7,167 in GOODS-N and 8,245 in GOODS-S. Galaxies in the {full} sample have a median S/N of about 4.0 in F090W and 13.2 in F277W. For each galaxy in the sample, its stellar mass and SFR are measured by fitting its combined HST+JWST SED (section \ref{sec:masses}). About 8\% (1,195/15,412) of the full sample has spectroscopic redshifts. The distribution of the full sample in terms of redshift and stellar mass is shown in Figure \ref{fig:sample_completeness}.

{To identify clumps within a mass-complete sample,} galaxies with stellar masses $\geq 10^8~M_{\odot}$ are selected. This is based on our completeness limits {(section \ref{sec:mass_completeness})}. In the highest redshift bin, $9<z<12$, this cut in mass is about 80\% complete. However, we choose to apply it across all redshift bins for consistency. Applying the mass cut removes 6,074 galaxies and leaves 9,338. 

Finally, galaxies are selected to be star-forming. This is done following pre-JWST studies, which studied star-forming clumpy galaxies \citep[e.g.,][]{Guo15, Shibuya16, HuertasCompany20, Sattari23}. In this work, galaxies with sSFR $\geq 10^{-9.5}~\textrm{yr}^{-1}$ are defined to be star-forming. Cutting on sSFR removes 217 galaxies. Our final sample contains 9,121 star-forming galaxies.

\begin{figure*}[t!]
    \centering
    \includegraphics[width=\textwidth, trim=0 0 3cm 0, clip]{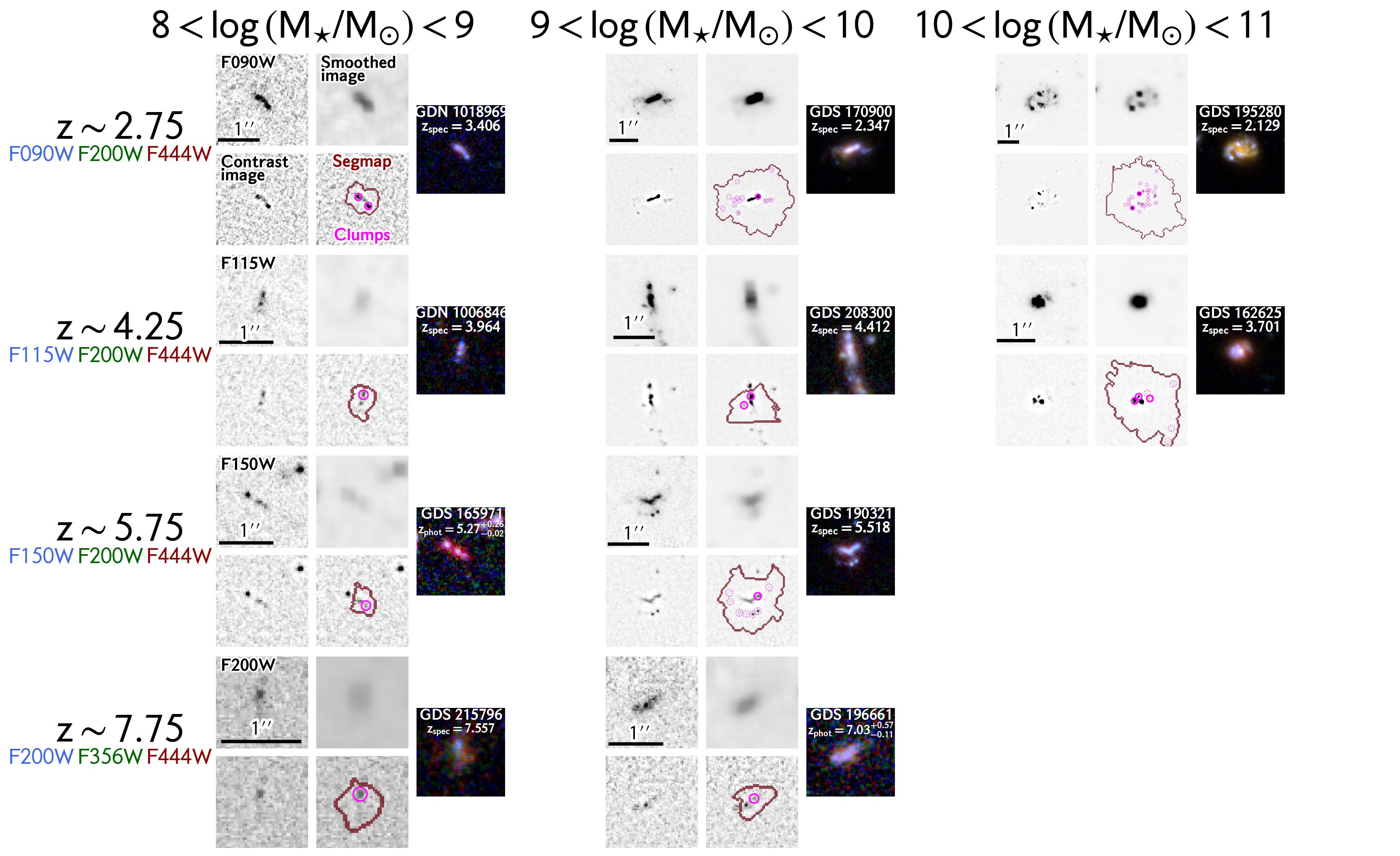}
    \caption{Example clumpy galaxies are shown {in each redshift (vertical) and stellar mass bin (horizontal), where a complete sample of clumps is detected. For each galaxy, images in the rest-frame NUV are shown to the left and an RGB image is on the right. Unmodified images are shown in the top left; smoothed images in the top right; contrast images in the bottom left, and detected clumps are circled in pink in the bottom right.} Clumps that meet the completeness limits listed in Table \ref{tab:completeness} are denoted with solid lines{, otherwise,} with dotted lines. The segmentation map of each galaxy is indicated with the red outline.}
    \label{fig:example_clumpy_galaxies}
\end{figure*}

\begin{figure*}[t!]
    \includegraphics[width=\textwidth]{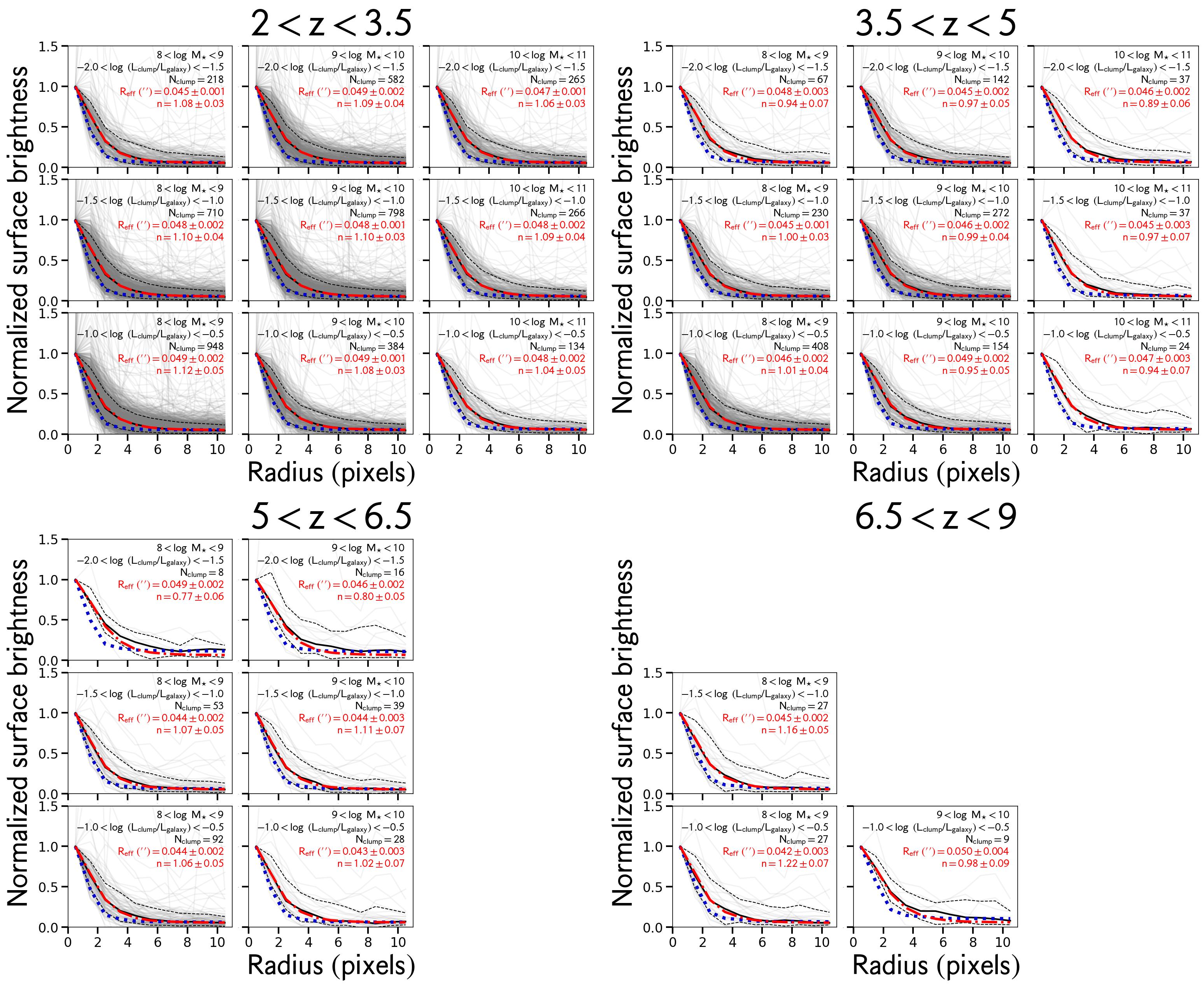}
    \caption{{Surface brightness profiles of detected clumps. In each panel, the surface brightness profiles of individual clumps are shown as thin gray lines. Those of the PSFs of the filters used to detect clumps are shown with dark blue dotted lines. Median profiles are plotted as thick, black, solid lines and running 16th and 84th percentiles are shown as lower and upper dashed lines. S\'{e}rsic profiles are fit to the median profiles and shown as red dashed-dotted lines. The effective radii and indices of the fits are written in red text. All surface brightness profiles are normalized to that of the innermost aperture used to measure them. Panels are arranged in blocks of 3x3 panels that are indicated with bins in redshift. Only panels with at least five detected clumps are shown. Within each block, host galaxy stellar mass increases from left to right and brightness of the clump compared with its host increases from top to bottom. The latter quantity is measured by assuming the clumps behave as point sources. In all panels, clumps are more extended than the PSFs. The angular sizes of clumps are nearly constant with redshift, but their physical sizes increase by a factor of $\sim2$ from $z\sim7.75$ to $z\sim2.75$.}}
    \label{fig:clump_light_profiles}
\end{figure*}

\section{Selection of Clumpy Galaxies}
\label{sec:selection_of_clumpy_galaxies}

\subsection{Identification of Clumps}
\label{sec:clump_identification}

Clumps are identified for all galaxies in the mass-selected, star-forming sample {as follows. For each galaxy in the sample, a 6\arcsec~by 6\arcsec~postage stamp is created in the bandpass corresponding to the rest-NUV. This image is then smoothed with a boxcar filter of width equal to 10 pixels. Our choice of filter width is justified with injection-recovery tests (section \ref{sec:completeness}). The smoothed image is subtracted from the original image to create a contrast image. Objects are then detected in the contrast image. This is done with the {\tt photutils} library. The $3\sigma$ clipped standard deviation of the background is measured and source detection is performed. Deblending is done with {\tt photutils} to separate overlapping clumps, with a minimum contrast of 0.001 and 32 thresholds. Objects are selected to be candidate clumps} if they span at least five contiguous pixels, each of which must be at least two standard deviations brighter than the background-subtracted image, and if the total S/N of the object is at least 3. {The minimum physical area of a candidate clump is 0.33 kpc$^2$ at $z=2$ and 0.06 kpc$^2$ at $z=12$.} {To be selected as clumps, candidates} must lie at least 0.\arcsec1 from the center and within the segmentation map created by the JADES team. This corresponds to a distance of 0.86 kpc at $z=2$ and 0.37 kpc at $z=12$. The former criterion is used to avoid confusion with a bulge or active galactic nucleus and the latter ensures that all detected clumps lie within a single galaxy and not within its neighbors projected on the sky. {Examples of the detection technique and identified clumps are shown in Figure \ref{fig:example_clumpy_galaxies}. The technique and selection criteria above broadly follow many pre-JWST studies \citep{Murata14, Guo15, Shibuya16, Soto17, Martin23, Sattari23}.}

\subsection{{Radial Surface Brightness Profiles and Aperture Photometry of Clumps}}
\label{sec:clump_profiles}

{In the HST data, clumps were largely found to be point sources \citep{Guo15, Shibuya16, Sattari23}. Given the higher spatial resolution of JWST, this assumption may no longer hold. The surface brightness profiles of clumps are shown in Figure \ref{fig:clump_light_profiles}, divided into bins of redshift, host galaxy stellar mass, and UV brightness of the clumps relative to their host galaxies. Profiles were measured with concentric circular apertures centered on the centroids determined by our clump detection routine. 

Clumps detected in this work have spatially extended surface brightness profiles, which are inconsistent with point sources. Empirical PSFs are measured and described in Appendix \ref{sec:psf}. Figure \ref{fig:clump_light_profiles} shows that in all bins, the median profiles are more extended than those of the PSFs.

To measure the flux of each clump, a typical surface brightness profile shape is first assumed. Then, the flux is directly measured within a small circular aperture. Background flux from the host galaxy is removed and an aperture correction is performed. This follows previous works \citep[e.g.,][]{Guo15, Shibuya16, Sattari23}, except for our treatment of the clumps as extended sources.

A S\'{e}rsic function is assumed for clump profiles, which is then convolved with the PSF and fit to the median observed profiles (Figure \ref{fig:clump_light_profiles}). The effective radii and S\'{e}rsic indices of the fits are estimated. Relative errors on the effective radii are generally small, of only a few percent. When fitting the median profiles, the background light from the host is not subtracted, but is included in the fit. We assume the light profile over $6-10$ pixels in radius is dominated by the background light, as most profiles flatten out at these radii (see also \citealt{Guo15}). 

{ The median clump profile changes little in angular size} with redshift, host galaxy mass, or clump brightness, { but exhibits an increase in physical size with decreasing redshift}. Over all bins shown in Figure \ref{fig:clump_light_profiles}, the median effective radius and S\'{e}rsic index from the fits are $R_{\rm{eff}} = $0.\arcsec046 and $n=1.04$. The 16--84th percentile range in $R_{\rm{eff}}$ is 0.\arcsec044 -- 0.\arcsec049 and that for $n$ is 0.94 -- 1.10. { The} physical size { increases} with redshift, such that the median effective radius of clumps grows from 0.23 kpc (0.\arcsec045) at $z\sim7.75$ to 0.39 kpc (0.\arcsec048) at $z\sim2.75$. { This is in excellent agreement with the typical sizes of clumps measured by \citet{Zhu26}, who examined the properties of clumps in JADES using different techniques from those used in this work. As argued by Zhu et al., since galaxies grew at a similar rate of a factor of $\sim2$ from $z\sim8$ to $z\sim3$ \citep{Allen25, Miller25}, the growth in clump sizes over these redshifts suggests that gas fragmentation occurred on small scales in the early Universe and grew with time.} { The clump sizes we measure are a} factor of a few larger than the median size of clumps within lensed galaxies, which was found to be 0.093 kpc over $3.5 < z <5.5$ \citep{Claeyssens25}. The larger sizes reported in this work may be due to our selection of unlensed galaxies. Individual clumps that may only be detectable in lensed galaxies could aggregate into larger ones in unlensed systems. 

Aperture photometry of the clumps is performed with {\tt photutils} as follows. First, the flux of a given clump is measured with a circular aperture of radius 4 pixels. Beyond this radius, the typical clump light profile is dominated by host galaxy light (see Figure \ref{fig:clump_light_profiles}). The background light is subtracted by measuring the flux in circular annuli over the radii 6-10 pixels. It is then divided by the area of the annulus to yield a mean surface brightness. This is multiplied by the area of the inner aperture with radius of 4 pixels to yield a background flux, which is then subtracted from the original flux measured. In each step, all other clumps are masked with circular masks of radius 4 pixels. The sizes of the apertures, annuli, and masks used here are also the same as those used in \citet{Guo15}. 

Last, aperture corrections are applied to background-subtracted fluxes of clumps. These are calculated by assuming clump light profiles broadly follow the median S\'{e}rsic profile $(R_{\rm{eff}} = $0.\arcsec046, $n = 1.04)$. This profile is then convolved with a PSF. The total flux of the convolved profile and that within a circular aperture of radius equal to 4 pixels are measured. The aperture correction is taken to be the ratio of the total flux to that within the aperture. Clump aperture fluxes are multiplied by factors of 1.511, 1.473, 1.477, 1.514, and 1.784 in GOODS-N and 1.533, 1.492, 1.509, 1.530, and 1.738 in GOODS-S for the F090W, F115W, F150W, F200W, and F277W bandpasses, respectively. These are determined using the empirical PSFs (Appendix \ref{sec:psf}). 
}

\begin{figure*}[t!]
    \includegraphics[width=\textwidth]{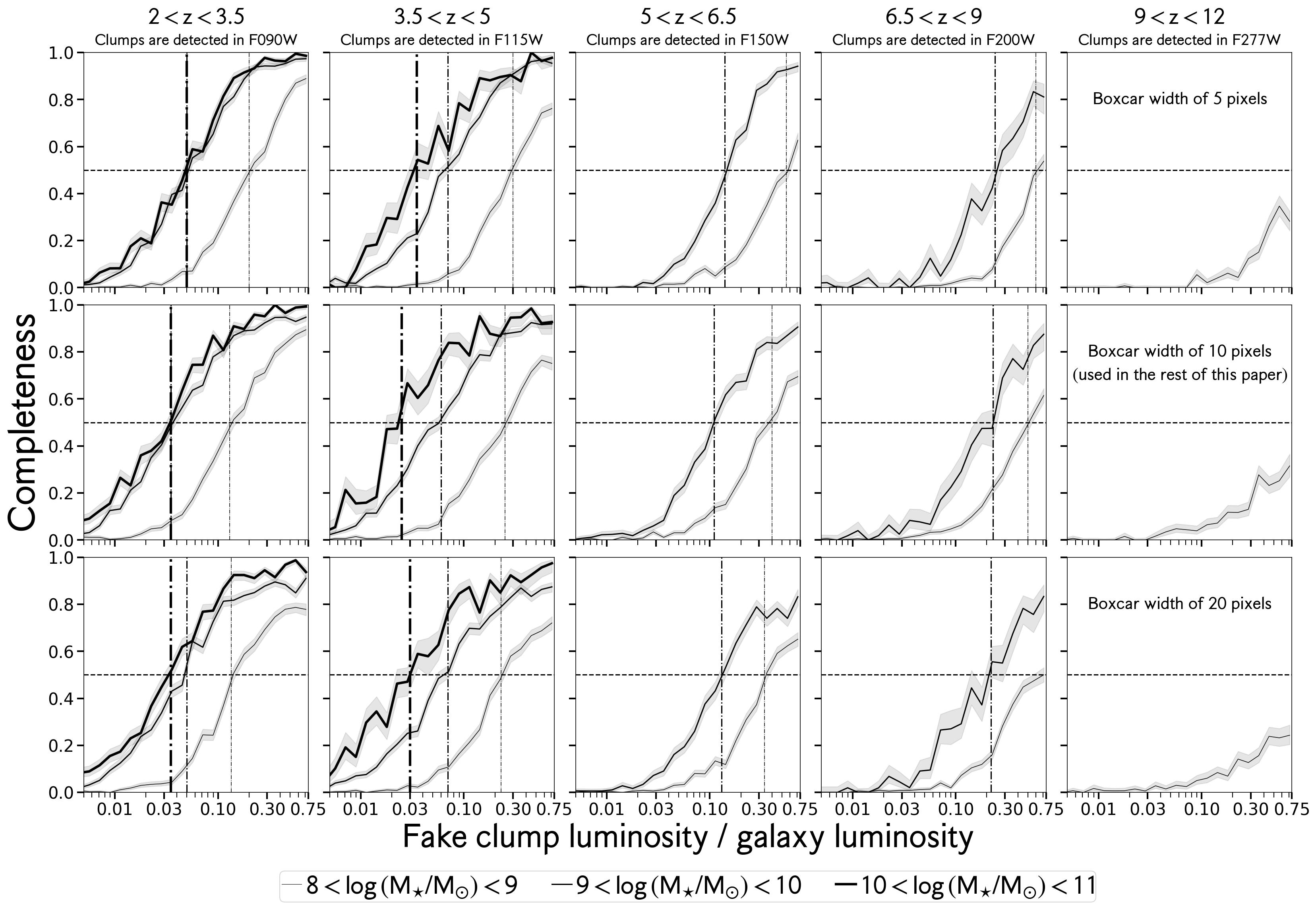}
    \caption{{Clump completeness functions in redshift (vertical) and stellar mass (horizontal) bins. The smoothing length used to detect clumps are also shown. Completeness is measured by injecting fake clumps to rest-frame NUV images of real galaxies and finding} the fraction of fake clumps recovered relative to the number inserted into each galaxy image. {The completeness is therefore measured as the ratio of fake clump luminosity to host galaxy luminosity above which $>$50\% of the fake clumps are detected.} A completeness of 50\% is denoted with the horizontal dashed lines. Minimum fractional luminosities that meet this completeness are indicated with vertical dashed-dotted lines. Those for more massive galaxies are denoted with thicker lines. Completeness curves are shown only for ranges in mass that include at least 10 galaxies. The error on the completeness is shown as the light shading around each line.}
    \label{fig:completeness_curves}
\end{figure*}

\subsection{Completeness of Clump Identification}
\label{sec:completeness}

To ensure that a complete sample of clumps is identified, we measure how faint clumps can be {in order for half of them to be detected} \citep{Guo15, Shibuya16, Sattari23}. This is done by adding fake clumps to images of randomly selected real galaxies, regardless whether they have clumps {or not}, and running the clump identification method described above. 

Fake clumps are modeled as {S\'{e}rsic profiles with the same effective radius and index as the median profile of the clumps. They are convolved with the empirical PSFs measured in this work. The PSF used to convolve the profile varies with redshift as described in section \ref{sec:sample_selection}.} Positions of the fake clumps are randomly selected to lie within the segmentation map of a given galaxy. Their brightnesses vary randomly between 0.1\% and 75\% of that of the host galaxy. In order to have sufficient statistics to determine the completeness, 100 galaxies are randomly selected in each of three stellar mass bins ($8<\log\left(M_{\star}/M_{\odot}\right)<9, 9<\log\left(M_{\star}/M_{\odot}\right)<10$, and $10<\log\left(M_{\star}/M_{\odot}\right)<11$) for each redshift bin. If fewer than 100 galaxies exist within a given bin in mass and redshift, all galaxies within that bin are selected. In total, 1,541 galaxies are randomly selected. A fake clump is added to the rest-frame NUV image of each galaxy. The clump identification algorithm is then run on the modified image to determine whether the fake clump is recovered. For each randomly selected galaxy, this procedure is repeated 50 times. The position and brightness of the fake clumps are randomly varied in each iteration. Uncertainties on the completeness are computed in a Bayesian fashion and assume a noninformative Jeffreys prior on the binomial distribution \citep{Cameron11}. 

{Since our clump detection procedure depends on the choice of the smoothing length of the boxcar filter, we perform the above procedure for three choices of smoothing length: 5, 10, and 20 pixels (0.\arcsec15, 0.\arcsec30, and 0.\arcsec60, respectively). If the smoothing length is too small, only the brightest clumps will be detected. If it is too large, spurious clumps may be detected, as noted by \citet{Guo15}. In this work, the optimal length is found to be 10 pixels (see below). It yields maximal completeness while being intermediate in length, thereby minimizing the chance of detecting spurious clumps. }

The {clump completeness functions are} shown in Figure \ref{fig:completeness_curves}. In each panel, the fraction of clumps detected by the clump identification algorithm is shown as a function of the fake clump luminosity. This is further shown as a function of stellar mass (lines of various thicknesses) and redshift (columns). {In each row, completeness curves are shown for a certain smoothing length that is used to detect clumps.} Clumps are more likely to be detected in higher-mass systems and lower redshifts. {Completeness limits are generally the highest for the shortest smoothing length of 5 pixels. They are smallest at the intermediate length of 10 pixels and change little for the longest smoothing length of 20 pixels. The intermediate length is chosen to be optimal. For all lengths considered, no clumps at $z\sim10.5$ are detected with at least 50\% completeness. In the following sections, we will only focus on the redshift range of $2<z<9$.}

Table \ref{tab:completeness} lists the minimum fractional luminosities as a function of stellar mass and redshift {for the optimal smoothing length of 10 pixels}. At $z\sim2.75$ and $\log \left(M_{\star}/M_{\odot}\right) > 9$, clump identification is 50\% complete at {about two} times fainter luminosities with JWST ({3.5}\%) than with HST (8\%; \citealt{Shibuya16}).

A galaxy is defined to be clumpy if it has at least one off-center clump with a rest-frame NUV luminosity that is greater than or equal to the completeness limits listed in Table \ref{tab:completeness}. These depend on stellar mass and redshift. 

\begin{deluxetable*}{l c c c}[t!]
\tablecaption{Minimum Fractional Luminosities Needed to Identify $\geq50\%$ of Clumps as a Function of Stellar Mass and Redshift\label{tab:completeness}}

\tablewidth{\textwidth}

\tablehead{Redshift Range & $8 < \log \left(M_{\star}/M_{\odot}\right) < 9$ & $9 < \log \left(M_{\star}/M_{\odot}\right) < 10$ & $10 < \log \left(M_{\star}/M_{\odot}\right) < 11$\\
& ($L_{\rm{clump}} / L_{\rm{host~galaxy}}$) & ($L_{\rm{clump}} / L_{\rm{host~galaxy}}$) & ($L_{\rm{clump}} / L_{\rm{host~galaxy}}$)}
\startdata
2--3.5 & {$\geq$13\%} & {$\geq$3.5\%} & {$\geq$3.5\%}\\
3.5--5 & {$\geq$25\%} & {$\geq$6\%} & {$\geq$2.5\%}\\
5--6.5 & {$\geq$40\%} & {$\geq$11\%} & $\ldots$\\
6.5--9 & {$\geq$50\%} & {$\geq$23\%} & $\ldots$\\
9--12 & $\ldots$ & $\ldots$ & $\ldots$\\
\hline
\enddata

\tablecomments{Each fraction is in terms of the rest-frame NUV luminosity of the host galaxy. Empty values indicate that either too few ($<10$) galaxies lie in a given mass and redshift bin or clumps could not be detected with at least 50\% completeness. {All minimum fractional luminosities here are reported for the optimal smoothing length of 10 pixels.}}

\end{deluxetable*}

\begin{figure*}[t!]
    \includegraphics[width=\textwidth]{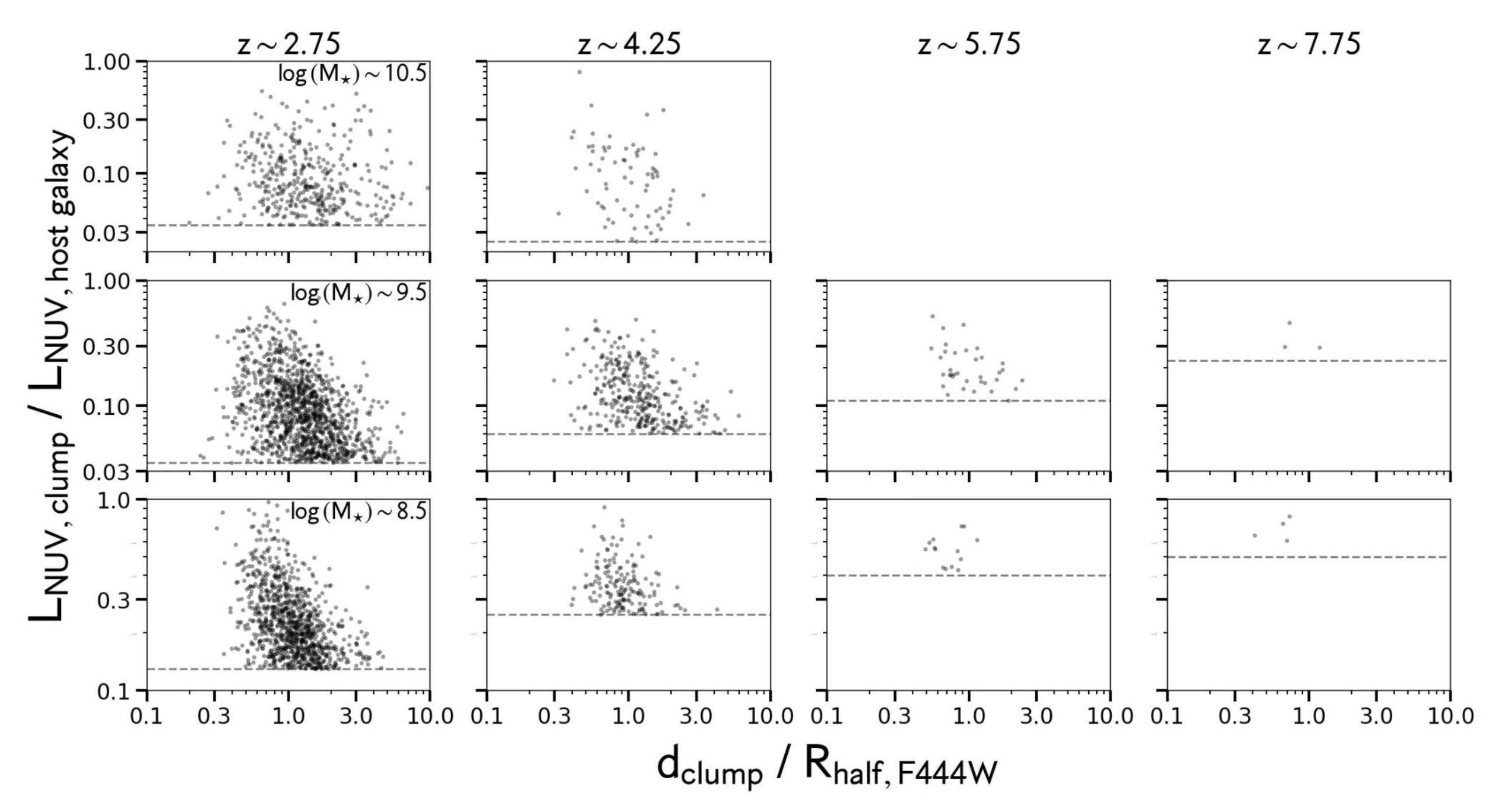}
    \caption{{ The ratio of clump to host galaxy NUV luminosity as a function of distance to the host center, normalized by the host half-light radius, in bins of redshift (columns) and host stellar mass (rows). Bins are shown only where there are at least 10 galaxies. In each panel, individual clumps are shown as small gray dots. Completeness limits are plotted as dashed lines. There is a radial trend, such that clumps are generally brighter as they approach the centers of their hosts, out to $z\sim5.75$. This generally agrees qualitatively with predictions for clumps that form {\it in situ} within galaxies \citep{Mandelker14, Mandelker17, Zanella19}. {\it Ex situ} clumps due to minor mergers are generally expected to lie in the outskirts (at $\gtrsim1 R_{\rm{half}}$) and be brighter than {\it in situ} ones (see the text). The majority of our sample of clumps is consistent with having formed {\it in situ}. }}
    \label{fig:clump_radial_trend}
\end{figure*}

\section{{ The Nature of the Detected Clumps}}
\label{sec:nature_clumps}

{ In this section, we discuss the physical characteristics of the clumps we find. In this work, clumps are taken to be compact, UV-bright regions, as was done in many previous studies. However, without additional information, such as physical properties from SED fitting or spatially resolved spectroscopy, it is difficult to determine whether clumps originated {\it in situ} within their host galaxies or {\it ex situ} from other processes, such as major or minor mergers \citep[e.g.,][]{Jiang13, Bowler17} or cold, inflowing gas that fragments into clumps \citep[e.g.,][]{Dekel09_theory, Ceverino10, Mandelker18, Kalita22}.}

{ One way to resolve this is with measurements of the masses, sizes, and/or ages of the clumps as a function of distance to the center of their hosts. As galactocentric distance decreases, {\it in situ} clumps are expected to be more massive and older \citep{Bournaud14, Mandelker14, Mandelker17}, which has also been observed over $0.5<z<5.5$ \citep{Shibuya16, Guo18, Claeyssens25, Kalita25}. {\it Ex situ} clumps caused by minor mergers are expected to be more massive (by factors of $\gtrsim 3$), older, and larger than {\it in situ} ones at fixed distance to the center, and lie predominantly in the outskirts of their hosts at $\gtrsim 1 R_{\rm{eff}}$ \citep[][]{Mandelker14, Mandelker17, Zanella19}. {\it Ex situ} clumps that develop via fragmentation of cold streams of gas are predicted to form in extended rings surrounding their hosts at radii of $0.1-0.3$ virial radii \citep{Danovich15, Mandelker18}. They are further expected to be more common at $z\gtrsim4$, as streams of gas are denser and cool faster at higher redshift.}

{ Measuring such physical properties, as listed above, is beyond the scope of this work. However, we attempt to determine which of the above scenarios best describes our sample of clumps with Figure \ref{fig:clump_radial_trend}. The figure shows the ratios of clump to host NUV luminosities against the distances of the clumps to the centers of their hosts, which are divided by the hosts' half-light radii in F444W, $R_{\rm{half, F444W}}$. The first quantity serves as a crude proxy for the mass ratio of a clump to its host. The second accounts for variations in host galaxy size. We find that the majority of the clumps are consistent with having formed {\it in situ} for three reasons.} 

{ First, there is a clear trend in Figure \ref{fig:clump_radial_trend} such that brighter (more massive) clumps are found closer to the centers of their hosts. From $\sim2R_{\rm{half, F444W}}$ to $\sim0.5R_{\rm{half, F444W}}$, clumps increase in brightness by factors of $\sim4-5$. This is similar to the expected growth in mass as a function of radius for {\it in situ} clumps \citep{Mandelker14, Mandelker17, Oklopcic17}. This is observed at all host masses at $z\lesssim4.25$ and there is a hint of it at $z\sim5.75$ at intermediate masses. This suggests an {\it in situ} origin for the clumps, as there is a continuity in the relative brightnesses of clumps across a broad range of host mass and size. If the clumps had largely formed through minor mergers, there should be a host mass dependence to the slope of this trend, as the minor merger fraction increases with increasing stellar mass, at least out to $z\sim1.5$ \citep{Ventou19}. As host mass increases, there is more scatter in the trend. This is likely due to a combination of greater amounts of dust in more massive galaxies \citep[e.g.,][and references therein]{Shapley23} and older clumps residing in more massive hosts \citep{Claeyssens25}.

Second, most of the clumps are located within $1-2 R_{\rm{half, F444W}}$ and are thus located within the bulk of their hosts' stellar light. {\it Ex situ} clumps from minor mergers are predicted to lie largely in the outskirts, beyond 1 $R_{\rm{half, F444W}}$ \citep{Mandelker17}. {\it Ex situ} clumps that formed via gas fragmentation would have formed at $\sim0.1-0.3$ virial radii, or roughly $3-10~R_{\rm{half, F444W}}$, assuming a typical ratio of half-light to virial radius of $\approx0.03$ \citep{Huang17}. This is well beyond where the bulk of the clumps lie. 

Last, clumps have typical sizes that are smaller than those of their hosts, consistent with expectations for {\it in situ} clumps \citep{Mandelker17, Oklopcic17, Zanella19}. We measure a median $R_{\rm{eff}}$ of the clumps to be 0.23 kpc at $z\sim7.75$ and 0.39 kpc at $z\sim2.75$ (section \ref{sec:clump_profiles}). At the lowest masses considered, $\log\left(M_{\star}/M_{\odot}\right)=8$, galaxies have typical $R_{\rm{eff}}$ of $\sim0.3$ kpc at $z\sim8$ \citep{Miller25} and $\sim1$ kpc at $z\sim1.5$ \citep{Nedkova21}. Since the galaxy size--mass relation has a positive slope \citep[e.g.,][]{vdW14, Ward24, Allen25}, the typical sizes of the clumps found here are smaller than those of their hosts at all masses considered, consistent with having formed {\it in situ}. In contrast, {\it ex situ} clumps from minor mergers are expected to have sizes similar to those of their hosts \citep{Zanella19}.}

\begin{figure*}[t!]
    \centering
    \includegraphics[width=\textwidth]{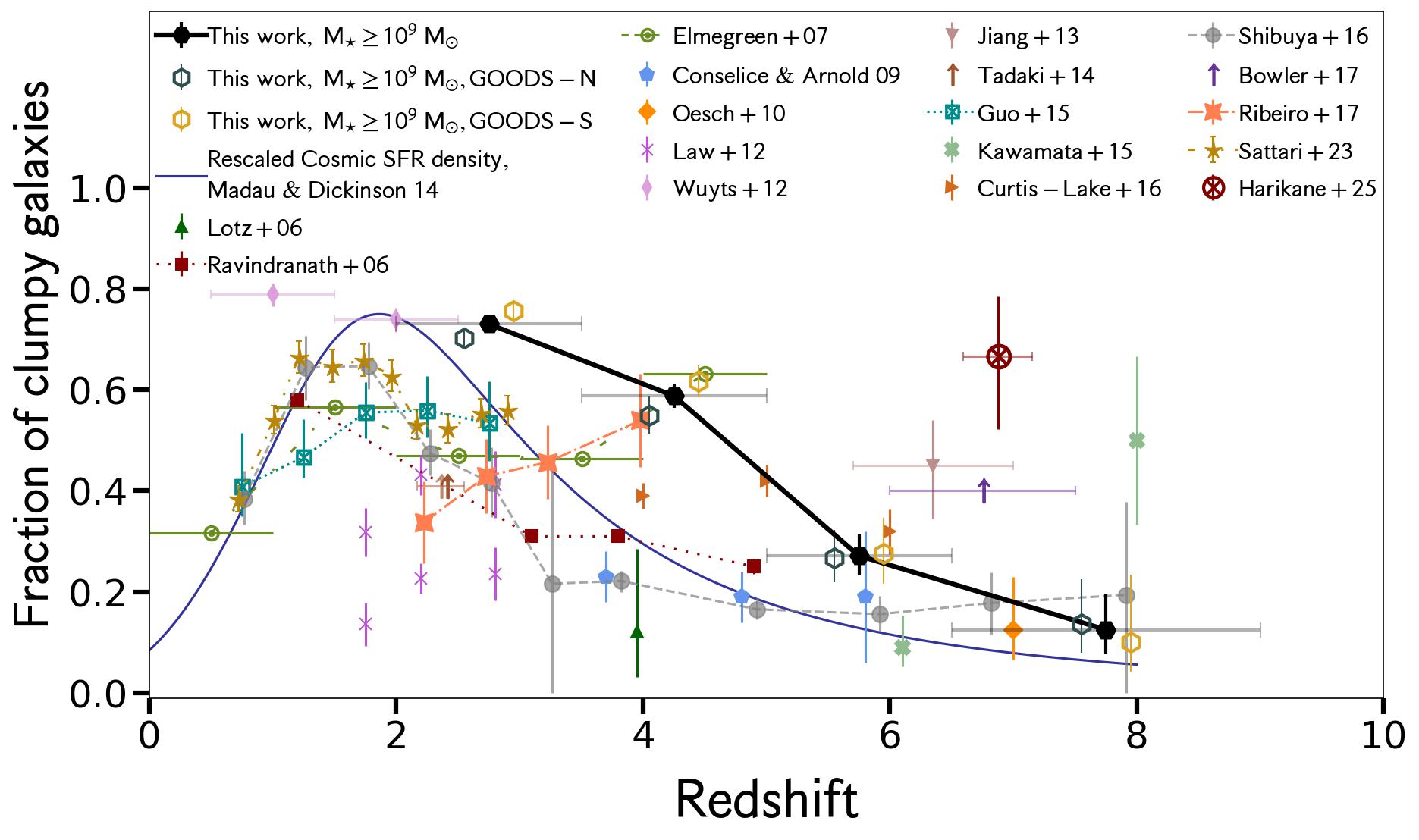}
    \caption{The fraction of clumpy galaxies, \fclumpy, over ${2<z<9}$. It {is highest} at $z\sim2.75$ and declines by a factor of $\sim3$ from $z\sim2.75$ to $z\sim5.75$. Galaxies are considered clumpy if they have at least one off-center clump. Measurements from this work for galaxies with stellar masses of at least $10^9~M_{\odot}$ are shown as black hexagons connected by a solid black line. \fclumpy~measurements for galaxies in GOODS-N {(GOODS-S)} are shown to the left {(right)} of the solid black line. The minimum fractional luminosity changes with redshift according to Table \ref{tab:completeness}. We compile measurements from the literature and plot them alongside ours.  The results {from these studies are largely consistent with each other}: \fclumpy~rises gradually from $z\sim8$ to $z\sim1.5-2$, when it peaks, and declines at lower redshifts. This is qualitatively consistent with the evolution of the cosmic SFR density, which is shown as a thin blue solid line and has been arbitrarily scaled vertically. The results of this study are consistent with the view that clumpy galaxies became more common at later times. However, {as redshift decreases, the \fclumpy~calculated in this work rises more steeply than that measured from the bulk of previous studies. At $z\sim2.75$ and $z\sim4.25$, our \fclumpy~measurements are higher than those of most studies by factors of $\sim1.3-3$.}}
    \label{fig:clumpy_fraction}
\end{figure*}

\section{Fraction of Clumpy Galaxies and Their Evolution with Redshift}
\label{sec:clumpy_fraction}

{For a galaxy to be clumpy, it must have at least one off-center clump identified in its rest-frame NUV image (section \ref{sec:clump_identification})}. {Each clump must have a minimum fractional luminosity relative to its host galaxy. These luminosities are} listed in Table \ref{tab:completeness}. The fraction of clumpy galaxies (\fclumpy) is defined to be the number of galaxies with clumps divided by the total number of galaxies in a given bin in {redshift and stellar mass}. All uncertainties are calculated assuming Bayesian confidence intervals on the binomial distribution, unless otherwise noted.

\begin{deluxetable*}{l c c c c c}[t!]
\tablecaption{Summary of Literature on Clumpy Galaxies\label{tab:literature}}

\tablewidth{\textwidth}

\tablehead{Reference & Sample $\left(N_{\rm{galaxy}}\right)$ & Stellar Mass Range & Redshift Range & Method & Detection Band\\
& & $\log\left(M_{\star} / M_{\odot}\right)$ & & &}
\startdata
\citet{Lotz06} & LBGs\tablenotemark{a}\tablenotemark{b} (82) & $\gtrsim9.5$ & $\sim4$ & $CAGM_{20}$\tablenotemark{c} & rest-frame UV\\
\citet{Ravindranath06} & SFGs\tablenotemark{d} (5719) & $\gtrsim9.6$ & 0.95--4.9 & Visual & rest-frame UV\\
\citet{Elmegreen07} & Starbursts (1003) & N/A & 0--5 & Visual & F775W\\
\citet{Conselice09} & LBGs (133) & $\gtrsim9$ & 4--6 & $CASGM_{20}$\tablenotemark{e} & rest-frame UV\\
\citet{Oesch10} & LBGs (21) & $\gtrsim8$ & 7--8 & Visual & rest-frame UV\\
\citet{Law12} & SFGs (309) & 9--11 & 1.5--3.6 & $CAGM_{20}$ & rest-frame UV\\
\citet{Wuyts12} & SFGs (649) & $>10$ & 0.5--2.5 & Algorithm & rest-frame 2800 \AA\\
\citet{Jiang13} & LBGs (24) & $\gtrsim9.2$ & 5.7--7 & $CAGM_{20}$ & rest-frame UV\\
\citet{Tadaki14} & H$\alpha$ emitters (100) & 9--11.5 & 2--2.5 & Algorithm & F606W and F160W\\
\citet{Guo15} & SFGs (3239) & 9--11.5 & 0.5--3 & Algorithm & rest-frame NUV\\
\citet{Kawamata15} & LBGs (39) & $\gtrsim8$ & 6--8 & Visual & rest-frame UV\\
\citet{Curtis-Lake16} & LBGs (1318) & $\gtrsim8.5$ & 4--8 & Asymmetry & rest-frame UV\\
\citet{Shibuya16} & SFGs and LBGs (13787) & 9--12 and $\gtrsim8.5$ & 0-8 & Algorithm & rest-frame UV and Optical\\
\citet{Bowler17} & LBGs (25) & $\gtrsim10$ & $\simeq7$ & Visual & rest-frame UV\\
\citet{Ribeiro17} & SFGs (1242) & 9.35--11.5 & 2--6 & Algorithm & F814W\\
\citet{HuertasCompany20} & SFGs (1575) & 9--12 & 1--3 & Algorithm & rest-frame UV and Optical\\
\citet{Sattari23} & SFGs (6767) & $>9.5$ & 0.5--3 & Algorithm & rest-frame 1600 \AA\\
\citet{Harikane25} & SFGs (12) & $\gtrsim8.5$ & 6.595--7.154 & Algorithm & F115W\\
This work & SFGs (9121) & $\geq8$ & 2--{9} & Algorithm & rest-frame NUV\\
\hline
\enddata

\tablecomments{$^a$Lyman-break Galaxies. $^b$Stellar mass ranges for samples comprising Lyman-break galaxies are estimated using the relations between absolute UV magnitude and stellar mass shown in Fig. 1 in \citet{Shibuya15}. $^c$``$C:$''concentration; ``$A:$''asymmetry; ``$G:$'' Gini coefficient; $M_{20}:$ second-order moment parameter. $^d$star-forming galaxies. $^e$``$S:$'' clumpiness. See \citet{Conselice14} for definitions of the $CASGM_{20}$ morphological parameters.}

\end{deluxetable*}

\subsection{Evolution with Redshift}
\label{sec:frac_redshift}
Figure \ref{fig:clumpy_fraction} shows \fclumpy~as a function of redshift. Measurements from this work for galaxies with stellar masses greater than $10^9~M_{\odot}$ are shown as black hexagons connected by a thick solid line. This limit in mass is constant with redshift and chosen {because} it is comparable to the mass limits used by other studies (see Table \ref{tab:literature}). 

The fraction of clumpy galaxies is highest ({$\sim70\%$}) at $z\sim2.75$ and decreases with increasing redshift to $\sim20$\% at {$z\sim7.75$}. {It changes the most from $\sim30\%$ at $z\sim5.75$ to $\sim60\%$ at $z\sim4.25$}. \fclumpy~measurements in either GOODS-N or GOODS-S agree with this trend {within $\leq3\%$}. 

Our \fclumpy~measurements are also compared with the cosmic SFR density (SFRD), as measured by \citet{MadauDickinson14}{, in Figure \ref{fig:clumpy_fraction}.  It has been rescaled along the y-axis to line up with the majority of the points in the figure.} The fraction of clumpy galaxies rise{s} with decreasing redshift more steeply than the SFRD. However, the net change in \fclumpy~from  ${z\sim7.75}$ to $z\sim2.75$, an increase of {almost} a factor of 4, is similar to that in the SFRD over the same range in redshift. This seemingly suggests a connection between the frequency of clumpy galaxies and the volume-averaged SFR {as a function of} redshift. 

\subsection{Comparison with Other Studies}
\label{sec:comparison_literature}

Our \fclumpy~measurements are also compared with those from other studies in Figure \ref{fig:clumpy_fraction}. The samples and methods of these are summarized in Table \ref{tab:literature}. The uncertainties on \fclumpy~were not published in some studies. They were calculated in this work assuming Bayesian binomial confidence intervals, wherever feasible. Details on the extraction of \fclumpy~measurements from earlier studies are provided {below.}

{Following \citet{Shibuya16}, we include \fclumpy~measurements from \citet{Lotz06, Ravindranath06, Conselice09, Oesch10, Law12, Jiang13, Kawamata15} and \citet{Curtis-Lake16}. These were calculated using the fractions of galaxies undergoing mergers and/or exhibiting disturbed morphologies or multiple cores. In \citet{Ravindranath06}, \citet{Oesch10}, and \citet{Kawamata15}, morphologies were determined via visual inspection. In the other works, morphological parameters were calculated.}

{Following \citet{Guo15}, we also include \fclumpy~measurements from \citet{Elmegreen07}. In the latter work, galaxies were visually inspected and categorized. \fclumpy~was inferred using the abundances of clump cluster, spiral, and elliptical galaxies. Other categories (chain galaxies, double nuclei, and tadpoles) were excluded as they generally have small axis ratios and would not have been included according to our sample selection (section \ref{sec:sample_selection}). }

{\fclumpy~from \citet{Bowler17} was obtained from their merger fraction, which was determined via visual inspection.}

For all other studies listed in Table \ref{tab:literature}, \fclumpy~measurements were given explicitly. {In \citet{Wuyts12}, \fclumpy~was measured by searching for galactic regions with blue colors and high surface brightnesses, based on visual inspection of coadded surface brightness profiles. Clumps in \citet{Tadaki14} were identified with the {\sc clumpfind} code \citep{Williams94}, which uses multiple isophotes to find clumps. In \citet{Guo15} and \citet{Shibuya16}, clumps are found using algorithms that are nearly identical to the one described in section \ref{sec:clump_identification}. \citet{Ribeiro17} detected clumps by identifying disconnected regions at various isophotal thresholds. \citet{HuertasCompany20} used neural networks to detect clumps with a training set of simulated images of galaxies. \citet{Sattari23} identified clumps by modeling the smooth, background light of host galaxies with Fourier transforms and subtracting this model to reveal clumps. Detection techniques similar to those used in \citet{Guo15} and \citet{Shibuya16} were then applied to the background-subtracted images. Last, clumps were detected in \citet{Harikane25} by performing source detection on images in the F115W bandpass. Subcomponents detected with $>5\sigma$ significance were then visually inspected.}

{For the rest of the studies shown in Figure \ref{fig:clumpy_fraction}, we find \fclumpy~to be $\sim30-40\%$ at $z\sim0.5$, peak at $\sim60-80\%$ at $z\sim2$, and decline to $\sim40-60\%$ at $z\sim3$ (although lower fractions of $\sim20-40\%$ were obtained by \citealt{Law12}). The decline continues to $\sim10-30\%$ at $z\sim4$ (although \citealt{Elmegreen07} and \citealt{Curtis-Lake16} measured higher fractions of $\sim40-50\%$). At $z\gtrsim4$, previous measurements of \fclumpy~largely divide into two groups: lower fractions of $\sim10-20\%$ that are approximately constant with redshift and higher fractions of $\sim30-70\%$ over $4<z<8$.}

The results of this work {are consistent with those of most previous studies at $z\gtrsim5.75$ and largely inconsistent at lower redshifts.} We find that, {as redshift decreases, \fclumpy~rises more steeply in this work than in the bulk of earlier studies. At $z\gtrsim5.75$, \fclumpy~as measured in this work is $\sim20\%$, similar to what was measured earlier at these redshifts. At $z\sim4.25$, we find \fclumpy~to be $\sim60\%$, whereas most earlier works found values between 20 and 40\%. At $z\sim2.75$, our measure of \fclumpy~is $\sim70\%$, while most other works found it to be between 40\% and 60\%.} 

\subsection{Comparing Clumpy Galaxy Fractions between the HST and JWST Data}
\label{sec:hst_jwst_comparison}

\begin{figure*}[t!]
    \includegraphics[width=\textwidth]{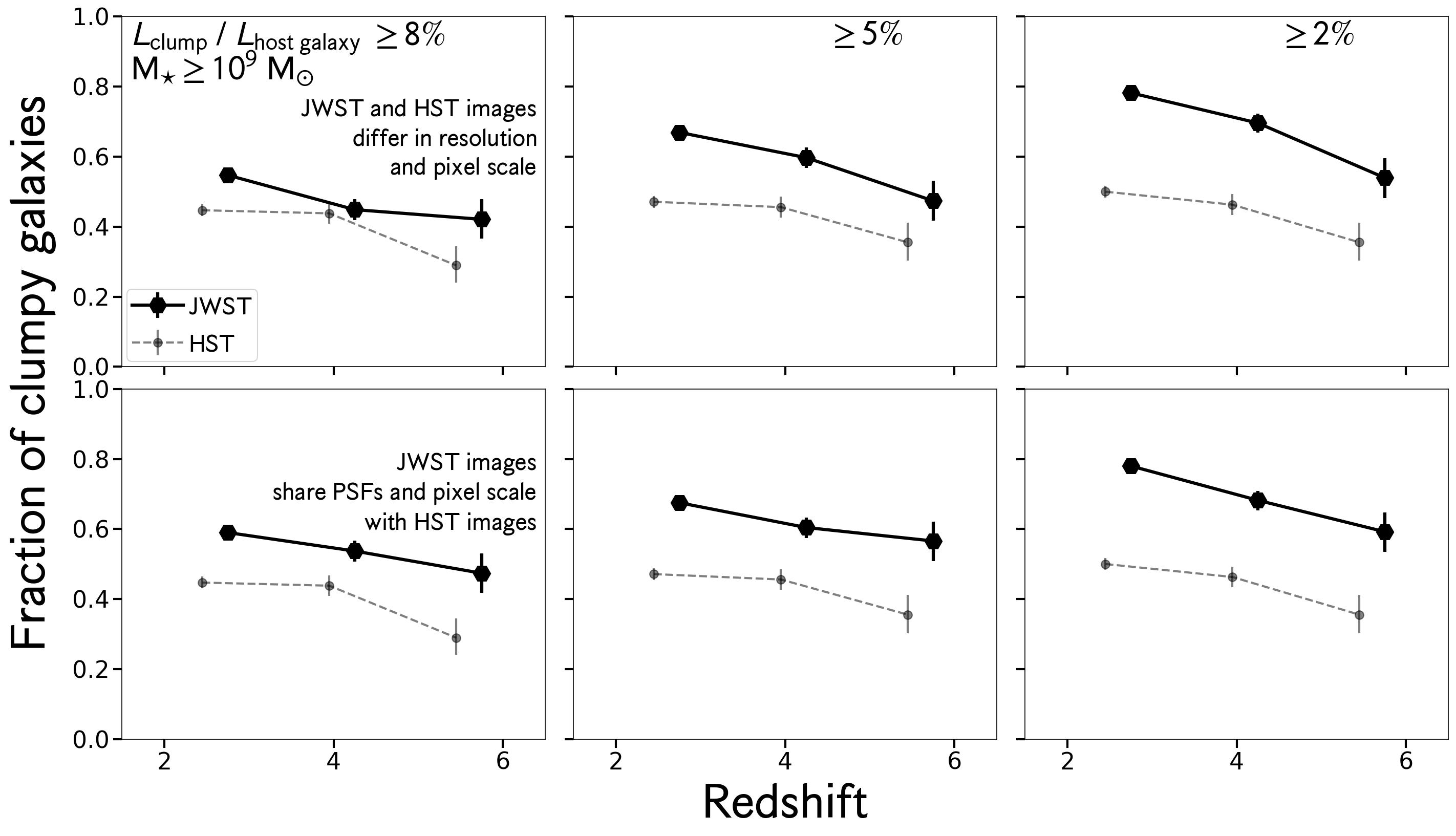}
    \caption{{To understand why \fclumpy~is higher with JWST data compared with previous data from HST, it is calculated with images from both telescopes as a function of redshift in each panel, minimum fractional luminosity (a proxy for image depth, columns) and whether JWST images are at their original pixel scale and resolutions (top row) or those of HST images (bottom). \fclumpy~from JWST data, as shown with the large hexagons and solid lines, lies systematically above that from HST data, plotted as small circles and dashed lines. As the minimum fractional luminosity gradually decreases, from left to right, \fclumpy~from JWST data progressively increases. Importantly, this is seen in both sets of JWST images, which are consistent with each other to within 9\%, at worst. However, \fclumpy~from HST images barely changes. Taken together, this suggests that the increased depth of JWST is the primary reason for the higher \fclumpy~measured in this work. Difference in resolution between HST and JWST plays a secondary role. Clumps are treated as point sources for only this comparison, as they were found to have light profiles consistent with point sources in work based on HST data.}}
    \label{fig:jwst_hst_comparison}
\end{figure*}

Our results on the redshift evolution of \fclumpy~disagree with those of most studies {listed in Table \ref{tab:literature} at $z\sim2.75$ and $z\sim4.25$.} All but {one} of these are based on {the} HST data, whereas this work is {entirely} based on JWST data. To determine the cause of the disagreement, we compute \fclumpy~using both HST and JWST images for a subset of our sample. {We measure \fclumpy~as a function of redshift at various minimum fractional luminosities. The aim is to assess to what extent the difference in resolution and sensitivity between the telescopes affect the measurement of \fclumpy. In the following, we demonstrate that} it is largely the difference in sensitivity that is responsible for the disagreement in the \fclumpy~measurements, {with difference in resolution having a secondary effect}. 

{To select galaxies for this comparison,} we match the mass-selected, star-forming sample (end of section \ref{sec:sample_selection}) with photometric catalogs from {GOODS-N \citep{Barro19} and GOODS-S \citep{Guo13}}. We use a match radius of 0.\arcsec5 and match HST/Wide Field Camera 3 (WFC3) F160W magnitudes {from these catalogs and those from the JADES} to within 0.5 mag. There are 2,007 matches in GOODS-N and 2,189 in GOODS-S. As studies based on HST data limited their analyses to galaxies with stellar masses $\gtrsim10^9~M_{\odot}$ (see Table \ref{tab:literature}), we only consider {this mass range}. This leaves a total of 1,245 galaxies (592 in GOODS-N and 653 in GOODS-S) for our comparison.

Clumps are identified in HST images using {the same algorithm and detection criteria described in section \ref{sec:clump_identification}}{, with one exception. Instead of using the segmentation map from JADES as the outer boundary in detecting clumps, as is done in section \ref{sec:clump_identification}, a limit is placed on the distance of a given clump to the center of its host to be less than $8\times R_{\textrm{half, F444W}}$, the half-light radius in F444W, following \citet{Shibuya16}. For the purposes of the comparison here, this outer boundary criterion is also applied to JWST images.} HST images have a pixel scale of 0.\arcsec06 pixel$^{-1}$. We note that the criterion on the minimum number of contiguous pixels that must be spanned by a candidate clump depends on the HST instrument used: it is five for ACS and seven for WFC3 images {\citep{Shibuya16}}. We search for clumps in HST data over $2<z<6.5$ to ensure that the coverage in rest-frame wavelength is as similar as possible between HST and JWST (see section \ref{sec:sample_selection}). Clumps are identified in HST bandpasses that span the rest-frame NUV at various redshifts {with ACS/F850LP ($2<z<3.5$); WFC3/F125W ($3.5<z<5$); and WFC3/F160W ($5<z<6.5$)}. Cutting on redshift removes 18 galaxies, with 1,227 remaining. We make sure that \fclumpy~is measured using the same galaxies for both HST and JWST images. Therefore, any difference in \fclumpy~is due to discrepancies in depth and resolution between HST and JWST images and not the selection of the sample.

{When measuring the brightness of clumps in the HST and JWST images, they are assumed to be point sources, instead of extended sources, as was found in section \ref{sec:clump_profiles}. This is done because clumps were found to largely resemble point sources in works based on HST data \citep{Guo15, Shibuya16, Sattari23}. To treat clumps found in the JWST images on equal footing, they are also considered point sources only for the purpose of the comparison presented here. Details are given in Appendix \ref{sec:clumps_point_sources} on the completeness of detecting clumps when treated as point sources and the aperture corrections needed.}

{JWST images are matched to the resolution of HST images using empirical PSFs in the HST bandpasses adopted from the 3D-HST survey \citep{Skelton14}. They have a common pixel scale of 0.\arcsec06. They are upscaled by a factor of 2 to match the pixel scale of the JWST images. PSF-matching kernels are then created with {\tt pypher} \citep{Boucaud16}. After matching the resolution of HST and JWST, we performed photometry on the clumps in the same way as described in section \ref{sec:clump_profiles}.}

{Aperture corrections for the HST images are 1.234, 1.337, 1.447 in GOODS-N and 1.225, 1.334, and 1.436 in GOODS-S for the F850LP, F125W, and F160W bandpasses, respectively. When matching JWST PSFs to those from HST, their aperture corrections agree to within 1\%. When treating clumps as point sources in the JWST images, aperture corrections are 1.348, 1.307, and 1.299 in GOODS-N and 1.365, 1.328, and 1.330 in GOODS-S for the F090W, F115W, and F150W bandpasses, respectively.}

{Comparison of evolutionary trends for \fclumpy~between HST and JWST is presented in Figure \ref{fig:jwst_hst_comparison}. The minimum fractional luminosity used to identify clumps is varied in each panel. It is highest in the left panel, corresponding to 8\% of the host luminosity \citep{Guo15, Shibuya16}. It is lowered to 5\% in the middle panel and 2\% in the right panel. The latter is the lowest clump minimum fractional luminosity at which a complete sample of clumps can be detected in the JWST images, when treating them as point sources. \fclumpy~measurements from JWST that are calculated with images at the original resolution are shown in the top row, while those at the resolution of HST are shown on the bottom.}

{Two conclusions may be drawn from Figure \ref{fig:jwst_hst_comparison}. First, at fixed minimum fractional luminosity, \fclumpy~from JWST data changes little with resolution. Second, as the minimum fractional luminosity is progressively lowered, \fclumpy~from HST barely changes, while that from JWST gradually changes.}

{To assess how much overlap there is in the clumps detected between the HST and JWST datasets, we count the number of clumps in the HST images with minimum fractional luminosity $\geq 8\%$. Next, for each clump detected with HST, we compare its $x$ and $y$ coordinates with those of clumps detected in JWST images with the same pixel scale as HST for the same host galaxy. We then count the number of clumps in the JWST images that lie within a given separation in pixels from its counterpart in the HST images. At a separation of 1.5 pixels, which corresponds to the PSF FWHM of F850LP \citep[0.\arcsec09,][]{Guo13}, 49\% of clumps are matched between the two datasets. At a larger separation of 3 pixels, which is equal to the FWHM of the F160W PSF (0.\arcsec18), 72\% of the clumps are matched. There are a few sources of uncertainty in these measurements. The center of the galaxy is determined separately for the HST and JWST data and may not be identical between the two. Detection images from JWST are significantly deeper than the HST, which could affect the determination of the centers of galaxies. Similarly, since a clump must possess sufficient contrast with the local background to be detected, differences in depth will affect how well clumps are detected and deblended. This is evident in our measurements. There are about 2.5 times more clumps in the convolved JWST images than in the HST (2,512 vs. 989) for the same galaxies and at a minimum fractional luminosity of $\geq8\%$.}

\begin{figure*}[t!]
    \centering
    \includegraphics[width=\textwidth]{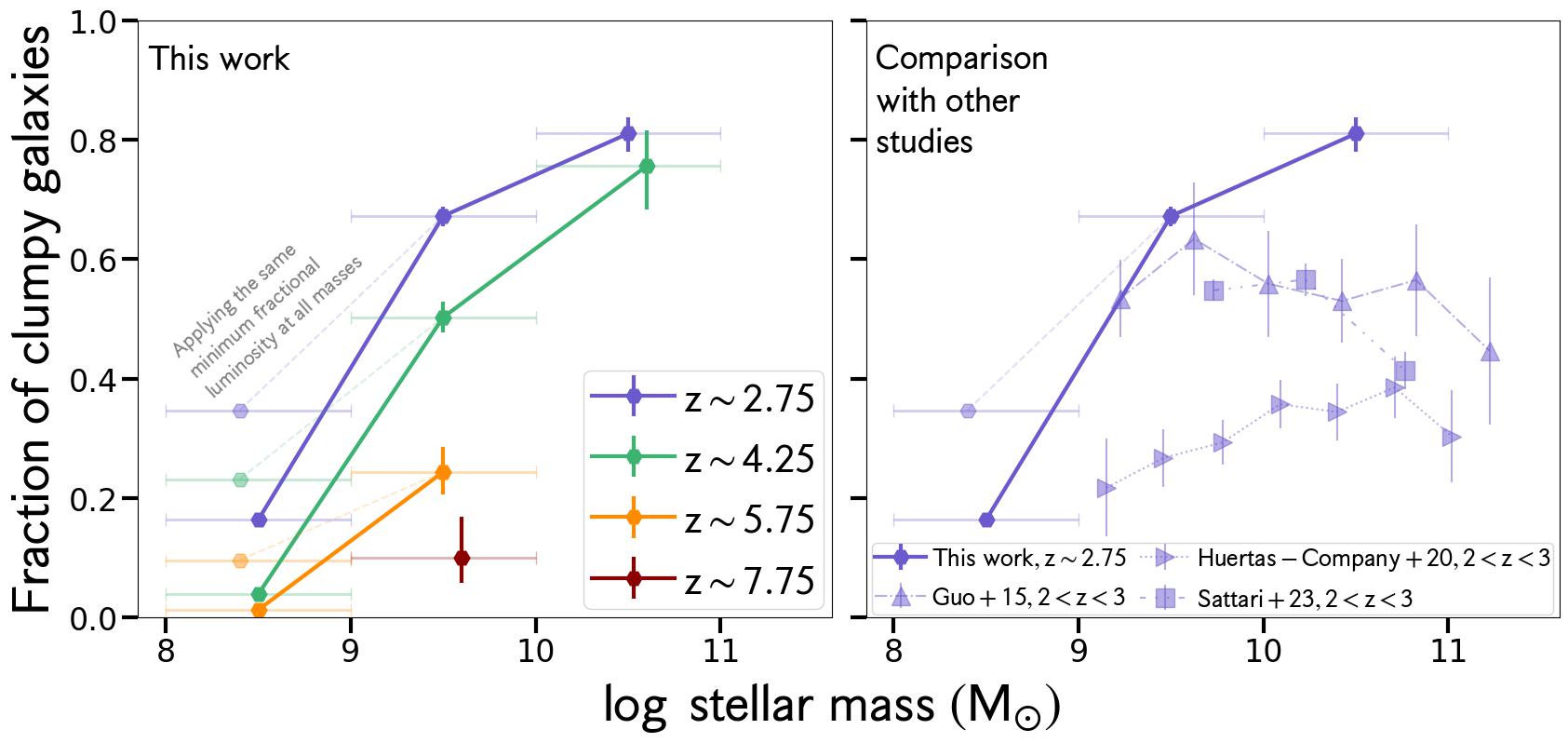}
    \caption{\fclumpy~as a function of host galaxy stellar mass using measurements from this work (left panel). Our measurements are shown in various colors for various redshift bins. The clump minimum fractional luminosities listed in Table \ref{tab:completeness} are used to compute \fclumpy. Because they vary with stellar mass, \fclumpy~is measured again for the lowest-mass galaxies, this time using the same as those for higher-mass galaxies. These alternative measurements are shown as fainter points connected with dashed lines. The present study finds a positive correlation between stellar mass and the fraction of clumpy galaxies, \fclumpy, over $2<z<12$. In the right panel, our \fclumpy~measurements at $z\sim2.75$ are compared with those from pre-JWST studies over $2<z<3$, plotted as various symbols. Two studies (\citealt{Guo15} and \citealt{Sattari23}) found an anticorrelation between stellar mass and \fclumpy. Another \citep{HuertasCompany20} found a positive correlation, but lower \fclumpy~values than ours by about a factor of 3.}
    \label{fig:clumpy_fraction_mass}
\end{figure*}

We attribute the results discussed above {mostly} to the difference in sensitivity between HST and JWST. The latter can detect point sources that are $\sim2-5$ times fainter than the former, depending on the wavelength\footnote{\href{https://jwst-docs.stsci.edu/jwst-opportunities-and-policies/jwst-general-science-policies/guidelines-for-proposals-where-jwst-and-hst-overlap-in-capabilities\#gsc.tab=0}{https://jwst-docs.stsci.edu/jwst-opportunities-and-policies/jwst-general-science-policies/guidelines-for-proposals-where-jwst-and-hst-overlap-in-capabilities\#gsc.tab=0}}. It is therefore not surprising that \fclumpy~is systematically higher when measured with JWST, when the faintest clumps are considered. {Furthermore, our finding that resolution has a secondary ($\leq$9 \%) effect on \fclumpy~is similar to what \citet{Guo12} found with HST images for a sample of 40 bright clumps in star-forming galaxies over $1.5<z<2.5$ and with stellar masses $M_{\star} \gtrsim 10^{10}M_{\odot}$.}

\section{{Clumpy Fraction as a Function of Stellar Mass}}
\label{sec:clumpy_fraction_mass}

The fraction of clumpy galaxies as a function of stellar mass of their host galaxy {is plotted in redshift bins in Figure \ref{fig:clumpy_fraction_mass}, left panel.} The fraction is calculated in three bins, spanning stellar masses $8<\log\left(M_{\star}/M_{\odot}\right)<9$, $9<\log\left(M_{\star}/M_{\odot}\right)<10$, and $10<\log\left(M_{\star}/M_{\odot}\right)<11$. Fractions are only computed for bins that have at least 10 galaxies.

At all redshifts, \fclumpy~increases with the stellar mass. At {a} fixed mass, as redshift increases, \fclumpy~decreases. At the lowest redshifts, $z\sim2.75$, \fclumpy~increases from ${\sim20\%}$ at $\log\left(M_{\star}/M_{\odot}\right)\sim8.5$ to $\sim70\%$ at $\log\left(M_{\star}/M_{\odot}\right)\sim9.5$ and $\sim85\%$ at $\log\left(M_{\star}/M_{\odot}\right)\sim10.5$. {At higher redshifts, \fclumpy~increases with stellar mass at a rate similar to that at $z\sim2.75$. At $z\sim4.75$, it grows from $\sim5\%$ at the lowest masses to $\sim70\%$ at the highest masses. At $z\sim5.75$, it rises from $\sim1\%$ at $\log\left(M_{\star}/M_{\odot}\right)\sim8.5$ to $\sim20\%$ at $\log\left(M_{\star}/M_{\odot}\right)\sim9.5$. At $z\sim6.75$, a complete sample of clumps is only detected at $\log\left(M_{\star}/M_{\odot}\right)\sim9.5$, for which \fclumpy~is $\sim17\%$.}

The above results were found by selecting clumps using the minimum fractional luminosities listed in Table \ref{tab:completeness}, which vary with stellar mass and redshift. However, the correlation between stellar mass and \fclumpy~still holds when selecting clumps with the same minimum fractional luminosity at all masses. The fraction of clumpy galaxies is calculated again for galaxies with stellar mass $\log\left(M_{\star}/M_{\odot}\right)\sim8.5$, with the minimum fractional luminosities this time being the same as those applied to higher-mass galaxies. {The result is shown in the left panel in Figure \ref{fig:clumpy_fraction_mass}.} The effect of lowering the minimum fractional luminosity for the lowest-mass galaxies is an increase in \fclumpy~by $\sim10-20\%$ at fixed redshift.

We note that the correlation observed in the present work may result from selection effects. Clumps are identified based on their brightnesses relative to their hosts. Therefore, clumps detected in more massive galaxies must be brighter, and hence, easier to detect, than those in less massive galaxies. Indeed, recent work on lensed galaxies has found that more massive galaxies contain more massive clumps \citep{Claeyssens25}. The physical properties of clumps that are identified in this work, such as stellar masses and SFRs, are needed to determine the impact of these selection effects.

{In Figure \ref{fig:clumpy_fraction_mass}, right panel,} we compare the correlation between stellar mass and \fclumpy~at $z\sim2.75$ with that measured by other studies over $2<z<3$. These measurements are shown as various symbols. \citet{Guo15} and \citet{Sattari23} found that, as stellar mass increases, \fclumpy~decreases. It was found to be $\sim60\%$ at $\log\left(M_{\star}/M_{\odot}\right)\sim9.5$ and $\sim40\%$ at $\log\left(M_{\star}/M_{\odot}\right)\sim11$. Both studies used HST data covering the rest-frame UV and developed detection algorithms to search for clumps. In contrast, \citet{HuertasCompany20} found that {it has a positive slope, with \fclumpy~increasing from $\sim20\%$ to $\sim30\%$ with as stellar mass increases from $\log\left(M_{\star}/M_{\odot}\right)\sim9$ to 11.}

Results from the present work at $z\sim2.75$ disagree with those from other studies at similar redshifts{, finding a} positive correlation between stellar mass and \fclumpy.

\section{Comparison with Hydrodynamic Simulations}
\label{sec:discuss_sims}

Here we discuss the implication of our results to galaxy formation scenarios. There are two key results. First, the fraction of clumpy galaxies, \fclumpy, increases with decreasing redshift over ${2<z<9}$. For galaxies with stellar mass $M{\star} \geq 10^9 M_{\odot}$, \fclumpy~is ${\sim10\%}$ at ${z\sim7.75}$ and ${\sim70\%}$ at $z\sim2.75$. {We find relatively higher \fclumpy~in JWST-selected samples compared to those selected by HST. This is due to the higher sensitivity of the JWST observations (section \ref{sec:clumpy_fraction} and Figure \ref{fig:clumpy_fraction}).} Second, at a fixed redshift, as stellar mass increases, the fraction of clumpy galaxies increases {too} (section \ref{sec:clumpy_fraction_mass} and Figure \ref{fig:clumpy_fraction_mass}).

\begin{figure*}[t!]
    \centering
    \includegraphics[width=\textwidth]{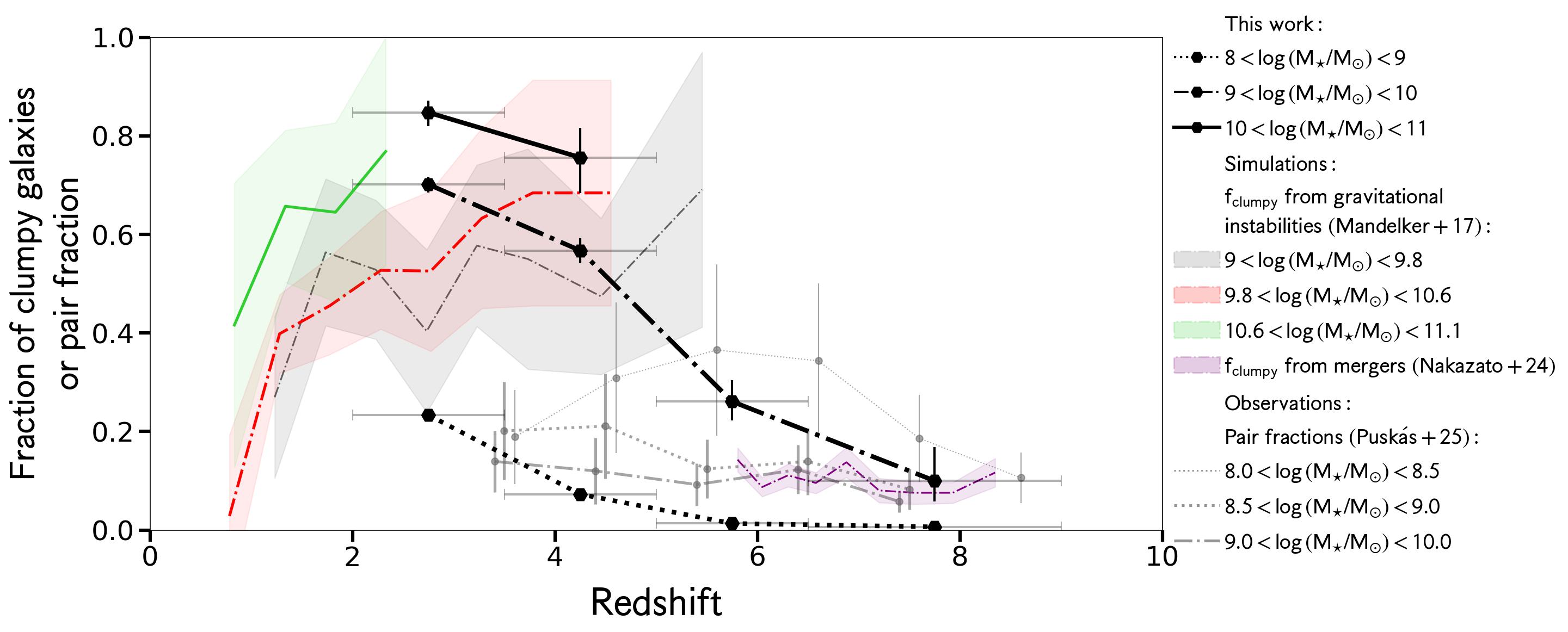}
    \caption{{Observed \fclumpy~values as a function of redshift in three stellar mass bins compared with predictions from simulations.} Uncertainties are shown as shaded regions. The fraction of clumpy galaxies that formed from violent disk instabilities \citep[VDI;][]{Mandelker17} at $z<6$ is shown as gray, red, and green shaded regions, in the order of increasing stellar mass. Predictions for clumpy galaxies that formed from major mergers over $6\lesssim z \lesssim 9$ \citep{Nakazato24} are shown as a purple line and shaded region. Observed pair fractions {as a function of stellar mass and redshift are shown as thin gray lines} \citep{Puskas25}. The results from this work are consistent {with \fclumpy~from simulations based on VDI and those based on major mergers at redshifts and masses where they overlap.} The observed pair fractions at $\log\left(M_{\star}/M_{\odot}\right) \leq 9$ are also consistent with the observed \fclumpy~measured here at similar masses and redshifts. }
    \label{fig:clumpy_fraction_vdi_mergers}
\end{figure*}

In this section, {we compare our results with simulations of clumpy galaxies. This is done to identify potential physical mechanisms for their formation. We compare with} zoom-in hydrodynamic cosmological simulations {from} \citet{Mandelker17} and \citet{Nakazato24}. The former studied 34 galaxies over $1 \lesssim z \lesssim 6$ from the VELA simulations \citep{Ceverino14, Zolotov15}, which had stellar masses of $\log\left(M_{\star}/M_{\odot}\right) \gtrsim 9$ at $z=2${, with clumps largely forming} through VDI. The latter analyzed 62 galaxies over $5.5<z<9.5$ from the FirstLight simulations \citep{Ceverino17}, which had stellar masses in the range $(0.5-6)\times10^{10}M_{\odot}$ at $z=5$. Most clumps in these simulations formed from major mergers. The \fclumpy~{estimates and predictions are compared in Figure \ref{fig:clumpy_fraction_vdi_mergers}.}

The \fclumpy~measurements are consistent with {predictions} from \citet{Mandelker17} at $z\lesssim5.75$ and $\log\left(M_{\star} / M_{\odot}\right) \geq 9$. The {predictions} increase rapidly with redshift from $\sim20\%$ at $z\sim1$ to $\sim50\%$ {at} $z\sim2$, above which they asymptote. They have uncertainties of $\sim20\%$, which result from Poisson noise due to a limited number of galaxies examined. {While our} \fclumpy~measurements peak at ${\sim70-80\%}$ at $z\sim2.75$ and decline with increasing redshift, {the model predictions do not change with redshift, when taking into account the uncertainties. This discrepancy is likely due to} differences in the way clumps are identified. \citet{Mandelker17} searched for clumps in projected density maps of their galaxies and computed \fclumpy~using those with stellar masses $>10^7~M_{\odot}$ {and} SFRs of at least 0.1\% those of their host galaxies' disks. In this work, clumps are detected in rest-frame NUV images and selected to be compact objects that are at least as bright as a fraction of their hosts' NUV luminosity, depending on redshift and stellar mass of the host galaxy. In both studies, clumps are selected to be off-center.

Our \fclumpy~measurements are also consistent with those from \citet{Nakazato24} at ${z\sim7.75}$ and ${\log\left(M_{\star} / M_{\odot}\right) \sim9.5}$. {At these redshifts and stellar masses, we find \fclumpy~to be $13$\%. A similar, near-constant value of $\sim10\%$ is obtained by \citet{Nakazato24} over $6<z<9$.} Galaxies studied {in} that work had stellar masses of $\lesssim10^9~M_{\odot}$ at $z\geq8.5$. Therefore, trends in \fclumpy~at the {intermediate} masses considered in this work are {consistent} with those from \citet{Nakazato24} at redshifts where {they} overlap. However, clumps were identified differently between these two {works}. In \citet{Nakazato24}, clumps were found in stellar mass surface density maps of young ($<10$ Myr) stars. Detectable clumps had stellar masses of $\gtrsim10^7~M_{\odot}$. This is {about an order of magnitude lower than the} faintest clumps that can be detected in this work at ${z\sim7.75}$ {and $\log\left(M_{\star} / M_{\odot}\right) \sim9.5$}, assuming clumps have similar NUV mass-to-light ratios as their hosts at these redshifts and stellar masses. {Furthermore,} in \citet{Nakazato24}, galaxies were defined to be clumpy if they had at least two clumps.

It is unclear from the simulations which processes dominate the formation of clumpy galaxies at other combinations of redshift and stellar mass, {particularly at the lowest masses}. Predictions for \fclumpy~for galaxies at these masses and redshifts are needed. Predictions from simulations that produce clumps through VDI are needed out to higher redshifts in order to distinguish clumps formed through VDI and those from mergers. Our \fclumpy~measurements at intermediate stellar masses ($9 < \log\left(M_{\star} / M_{\odot}\right) < 10$) are about 2 times higher than those from \citet{Nakazato24} at $z\sim5.75$. This suggests that major mergers may not significantly contribute to clumpy galaxy formation at these masses and redshifts (see below for support for this from observations). However, confirming this will require \fclumpy~measurements from simulations of clumpy galaxies that form via other mechanisms at these redshifts.

\section{Discussion}
\label{sec:discuss_obs}

{In this section, we discuss observations of the morphologies, gas kinematics, and pair fractions of star-forming galaxies with similar redshifts and stellar masses as those studied in this work. This is done to assess the plausibility of the mechanisms considered in the previous section, wherein our measurements of \fclumpy~were compared with predictions from simulations. The comparison suggests that} VDI and mergers {may be} the dominant mechanisms for clumpy galaxy formation, depending on redshift and stellar mass, in agreement with other studies \citep[e.g.,][]{Guo15, Shibuya16}. {The dominant mechanism for forming clumps appears to be VDI at $z\lesssim5.75$ and at intermediate-to-high stellar masses ($\log\left(M_{\star} / M_{\odot}\right) \geq 9$) and mergers at ${z\gtrsim7.75}$ and {intermediate} stellar masses (${\log\left(M_{\star} / M_{\odot}\right) \sim 9.5}$)}. We find that {the} observations {listed above} largely support the scenario described {here}. 

For clumps to form through VDI, their host galaxies must possess marginally stable disk components \citep[e.g.,][]{Bournaud07, Agertz09, Dekel09_theory, Ceverino10, Mandelker14}. star-forming galaxies over $2<z\lesssim6$ and $\log\left(M_{\star} / M_{\odot}\right) > 8$ have typical S\'{e}rsic indices of $n\simeq1$ \citep{vdW14, Shibuya15, Shibuya16, Nedkova21, Ormerod24, Sun24, Varadaraj24, Ward24, Miller25, Yang25}, which suggests they have disk-like structures. At higher redshifts, S\'{e}rsic indices are generally larger, which implies disk-like structure becomes less common (\citealt{Sun24, Miller25}; see also \citealt{Ferreira23, Kartaltepe23}). 

The stability of gaseous galaxy disks against fragmentation due to gravity is often quantified with the ratio of the circular velocity to velocity dispersion, $V/\sigma$ (see \citealt{vdK11} and \citealt{ForsterSchreiber20} for reviews). This is often interpreted to be the ratio of ordered to disordered motions \citep[e.g.,][]{Kassin07, Kassin12, ForsterSchrieber09, Epinat10, Stott16}. Galaxies are expected to be dominated by VDI when $V/\sigma \leq 2$ \citep{Dekel14, Zolotov15}. Studies that examined the gas kinematics of star-forming galaxies over $2 < z < 3.5$ found that those with stellar masses $\log\left(M_{\star} / M_{\odot}\right) >9$ have $V/\sigma \sim 0.5 - 4$, with higher values at lower redshifts and higher masses \citep{Gnerucci11, Simons17, Turner17, Wisnioski19}. VDI may therefore be the prominent process in the evolution of the majority of these systems. At higher redshifts ($4<z<8$), a wide range in $V/\sigma$ ($0.1 \lesssim V/\sigma \lesssim 20$) has been found in various studies \citep{Fraternali21, Rizzo21, Parlanti23, deGraaff24}. Recently, \citet{Danhaive25} modeled the H$\alpha$ kinematics of 272 galaxies over $z\approx3.9 - 6.5$ and $8<\log\left(M_{\star} / M_{\odot}\right)<10$ with the NIRCam grism. They measured an average $V/\sigma\approx2$, which suggests the majority of galaxies at these redshifts and stellar masses might be undergoing VDI. Large and deep surveys at high spatial and spectral resolution are needed to determine whether the majority of galaxies at even higher redshifts ($z>7$) could be experiencing VDI. 

For clumps to form mainly through mergers, it is expected that the fraction of clumpy galaxies and that of galaxies undergoing mergers are similar. The \fclumpy~measurements from the present study are compared with pair fractions measured by \citet{Puskas25} in Figure \ref{fig:clumpy_fraction_vdi_mergers}. In that work, the pair fraction was measured over $3<z<9$ and $8 < \log\left(M_{\star} / M_{\odot}\right) < 10$ in both of the JADES fields. \citet{Puskas25} computed the frequency of galaxies with major companions at projected separations of $5-30$ kpc and stellar masses of at least one-fourth those of primary galaxies. Their pair fractions agree with other recent measurements of the pair fraction at high redshift with JWST \citep{Dalmasso24, Duan25}.
{Our \fclumpy~measurements are consistent with pair fractions from \citet{Puskas25} at $z\sim7.75$ and over $9 < \log\left(M_{\star} / M_{\odot}\right) < 10$.} This suggests that major mergers are the dominant mechanism for clumpy galaxy formation {at $z\sim7.75$ and ${\log\left(M_{\star} / M_{\odot}\right) \sim 9.5}$}. At lower stellar masses, the pair fractions are {generally much higher than \fclumpy. This could suggest that clumps form inefficiently through mergers at these masses. However, given the lack of predictions in this mass regime, we leave interpretation to future work.}

At ${z<5.75}$ and ${\log\left(M_{\star} / M_{\odot}\right) \sim 9.5}$, {\fclumpy} is 3 to 4 times higher than the {pair fractions}. This implies that major mergers are not the dominant process through which clumps form at these masses and redshifts. This is also the case at the highest masses considered here. \citet{Duncan19} calculated pair fractions over $0.5 < z < 6$ and $\log\left(M_{\star} / M_{\odot}\right) > 9.7$ using HST data. The pair fraction at $\log\left(M_{\star} / M_{\odot}\right) > 10.3$ was found to be $\sim10-15\%$ over $2\lesssim z \lesssim4$ and increases to $\sim30-40\%$ at $z\sim6$. This is inconsistent with our \fclumpy~over $10<\log\left(M_{\star} / M_{\odot}\right) <11$, which decreases from ${81\%}$ at $z\sim2.75$ to ${76\%}$ at $z\sim4.25$. 

{We note that VDI and mergers are not the only mechanisms known to produce clumps. Another one, particularly active at high redshift, comes from turbulence in the interstellar medium driven by accretion of gas and stellar feedback \citep{Sun26}. This can lead to strong gas temperature and density fluctuations that result in spatially extended, clumpy star formation.}

\section{Conclusions}
\label{sec:conclusions}

We use JWST/NIRCam imaging from the JADES survey \citep{Eisenstein26, Eisenstein25, Rieke23, Bunker24, DEugenio25} to identify clumpy star-forming galaxies over ${2<z<9}$. The evolution of the fraction of clumpy galaxies, \fclumpy, with redshift and correlations with other physical properties were examined. A sample of 9,121 galaxies with stellar masses $\log\left(M_{\star} / M_{\odot}\right) \geq 8$ and $\log\left(\rm{sSFR/yr}^{-1}\right) \geq -9.5$ was selected by fitting their combined HST and JWST SEDs. Clumps were selected in a way similar to earlier studies based on HST \citep[e.g.,][]{Guo15, Guo18, Shibuya16, Sattari23}. The algorithm {described in section \ref{sec:clump_identification}} was used to find off-center clumps in rest-frame NUV images of galaxies. The completeness of the clump detection procedure was determined by injecting fake clumps into images of real galaxies. Clumps were selected to be $\geq 50\%$ complete as a function of redshift and stellar mass. 

Our main conclusions are summarized below. 
\begin{enumerate}
    \item We find that \fclumpy~increases with decreasing redshift from ${\sim10\%}$ at ${z\sim7.75}$ to ${\sim70\%}$ at $z\sim2.75$ (section \ref{sec:clumpy_fraction} and Figure \ref{fig:clumpy_fraction}). This is similar to the relative increase in the cosmic volume-averaged star-formation density over the same range in redshift. Compared with the bulk of pre-JWST studies, \fclumpy~from this work is higher at fixed redshift and increases more steeply with decreasing redshift. 
    \item The higher \fclumpy~found in this work is {mostly} due to the higher sensitivity of JWST compared with HST (section \ref{sec:hst_jwst_comparison} and Figure \ref{fig:jwst_hst_comparison}). We calculate \fclumpy~for a sample of 1,227 galaxies that are detected in both HST and JWST images and compare measurements from both sets of images. {We repeat this for another set of JWST images that share the resolutions of the HST images. \fclumpy~from JWST shows little change with resolution, but increases significantly when fainter clumps are selected.}
    \item We examine {the} correlation between \fclumpy~and {the stellar masses of their host galaxies}. There is a strong correlation between stellar mass and \fclumpy, such that more massive galaxies have higher \fclumpy~(section \ref{sec:clumpy_fraction_mass} and Figure \ref{fig:clumpy_fraction_mass}). This is the opposite trend of most previous studies.
    \item Our \fclumpy~measurements are consistent with a physical scenario in which clumps form largely through violent disk instabilities at $z\lesssim5.75$ and over $9 < \log\left(M_{\star} / M_{\odot}\right) < 11$ and through mergers at ${z\sim7.75}$ and over ${9 < \log\left(M_{\star} / M_{\odot}\right) < 10}$ (section \ref{sec:discuss_sims} and Figure \ref{fig:clumpy_fraction_vdi_mergers}). This is done by comparing \fclumpy~measured from simulations with the observations. This physical scenario is bolstered when considering the observed morphologies, gas kinematics, and pair fractions of star-forming galaxies at redshifts and stellar masses that overlap with the sample examined in this work (section \ref{sec:discuss_obs}). 
\end{enumerate}

We release a catalog of physical properties derived from SED fits, the empirical PSFs, and code to reproduce Figure \ref{fig:clumpy_fraction} on Zenodo under an open-source Creative Commons Attribution license: \dataset[doi:10.5281/zenodo.19444334]{https://doi.org/10.5281/zenodo.19444334}.

\begin{acknowledgments}
{We thank the anonymous referee for constructive comments that improved the paper.}

{A.d.l.V. thanks Guochao Sun for enlightening discussions on formation of clumps in turbulent, high-redshift galaxies. A.d.l.V. also thanks Mark J. Achenbach for hitting the ``submit'' button on arXiv while A.d.l.V was driving in Hilo.}

We thank the JADES team for designing and preparing their observations and releasing a rich public dataset. We also thank the STScI staff for enabling this science. 

This research has made use of the SVO Filter Profile Service ``Carlos Rodrigo", funded by MCIN/AEI/10.13039/501100011033/ through grant PID2023-146210NB-I00. 

A.d.l.V. was partially supported by the National Science Foundation through grant AST-2319553.
\end{acknowledgments}

All of the JWST data used in this paper can be found in {\it MAST} \citep{Rieke23_doi, Williams23_doi}, as can all of the HST data used in this paper \citep{Faber11_doi}. 

\software{astropy \citep{2013A&A...558A..33A,2018AJ....156..123A,Astropy22}, 
{LMFIT} \citep{Newville20},
matplotlib \citep{Hunter:2007},
numpy \citep{harris2020array},
photutils \citep{Bradley20, Bradley23},
scipy \citep{2020SciPy-NMeth}
          }

\appendix

\section{Empirical Point-Spread Functions}
\label{sec:psf}

Empirical PSFs were measured using {\tt photutils} {\tt EPSFBuilder} routine, which is based on the technique developed by \citet{AndersonKing00}. We adopt the parameters used to construct PSFs in {\tt photutils} from \citet{ZhuangShen24}, who characterized JWST/NIRCam PSFs in Cycle 0 calibration programs. In the present work, PSFs were created using bright (F277W mag $< 22$), unsaturated, and isolated point sources that were visually inspected in each NIRCam bandpass. For the wide bandpasses, 16 point sources in GOODS-N and 17 in GOODS-S were used to estimate PSFs. For the medium bandpasses, six to ten point sources in either field were used, depending on the bandpass.

\begin{figure*}
    \includegraphics[width=\textwidth]{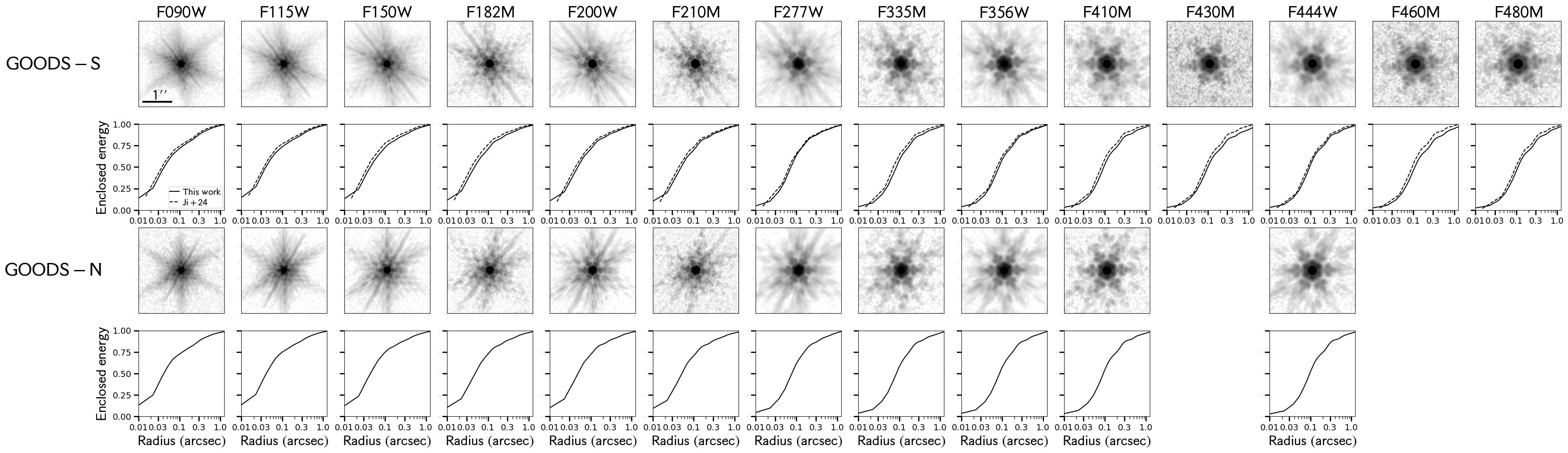}
    \caption{Empirical point-spread functions (PSFs) in 14 JWST/NIRCam bandpasses in JADES GOODS-N and GOODS-S are measured in this work and shown in this figure. In the top row are images of the PSFs in each bandpass in GOODS-S. Below it are the curves of enclosed energy as a function of radius for each PSF. This is repeated in the third and bottom rows for GOODS-N. The PSFs are estimated using the algorithm by \citet{AndersonKing00} and implemented in {\tt photutils} \citep{Bradley23}. The parameters that are used to construct the PSFs with {\tt photutils} are adopted from \citet{ZhuangShen24}. The curves of enclosed energy that are measured in this work agree to within 5\% with those measured by \citet{Ji24}, who also calculated empirical PSFs in the JADES GOODS-S field. The enclosed energy curves from that work are shown as dashed lines in the second row.  }
    \label{fig:psfs}
\end{figure*}

Empirical PSFs and their curves of enclosed energy are shown in Figure \ref{fig:psfs}. The latter are measured by summing the light from the PSFs in circular apertures at various radii and are plotted as solid lines in the Figure. At each radius, the value of the curve indicates the fraction of the total light enclosed. We compare the PSFs measured here with those from \citet{Ji24}, who calculated empirical PSFs in the JADES GOODS-S field. Their enclosed energy curves are shown as dashed lines in the second row. There is good agreement overall. At worst, the enclosed energy curves computed in this work for the F430M, F460M, and F480M bandpasses disagree with those from \citet{Ji24} by 5\% at most. For all other bandpasses, they disagree at most by about $2-3\%$.

\section{{Treating Clumps as Point Sources in JWST Images}}
\label{sec:clumps_point_sources}

{In this appendix, we briefly describe how clumps are identified in JWST images when they are considered to be point sources. While clumps in JWST data typically have extended light profiles (section \ref{sec:clump_profiles}), in order to compare \fclumpy~measurements from HST and JWST images (section \ref{sec:hst_jwst_comparison}), clumps must be treated as point sources. This is because they were largely found to have light profiles consistent with point sources in earlier work based on HST.}

\begin{figure*}
    \includegraphics[width=\textwidth]{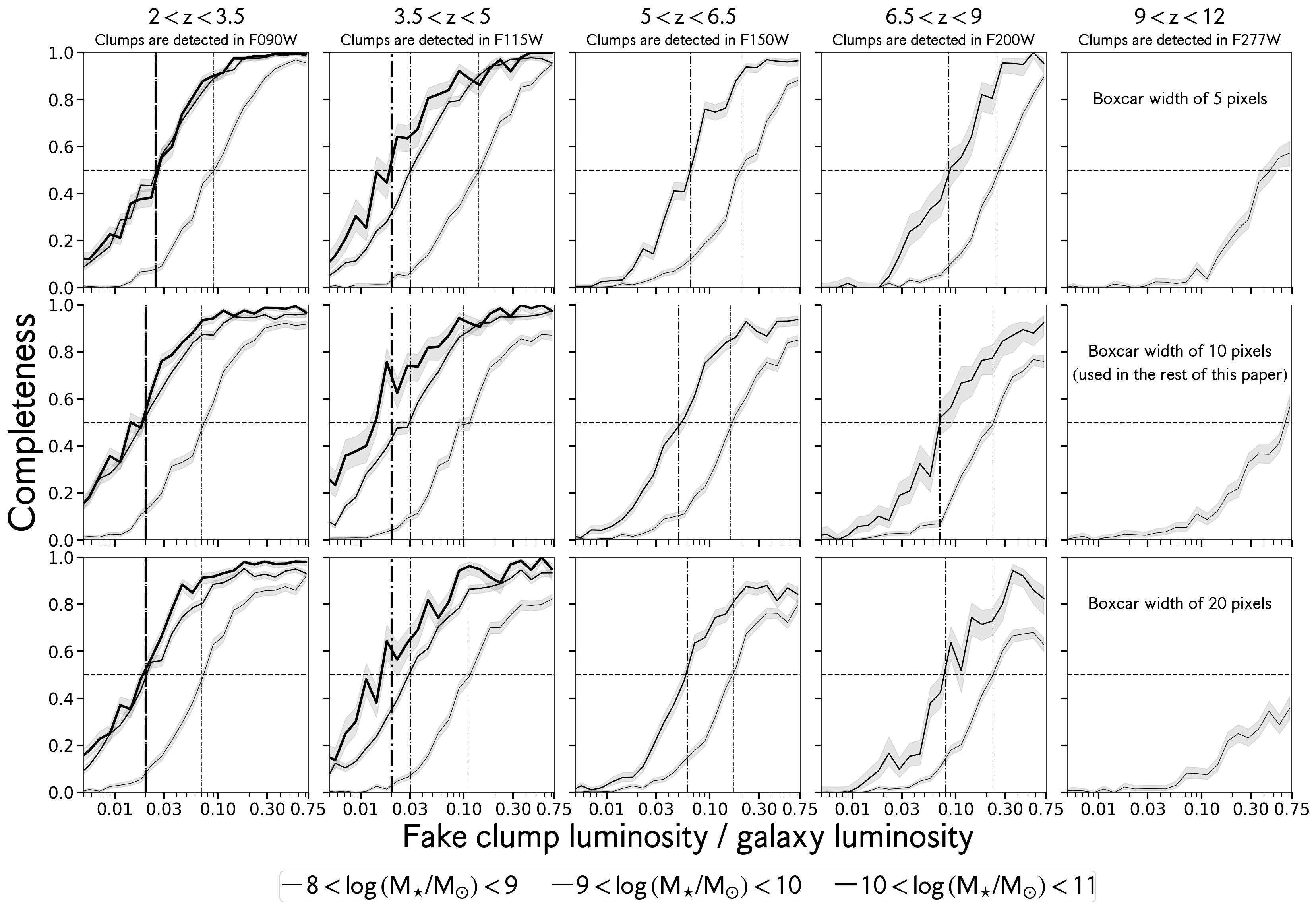}
    \caption{{Same as in Figure \ref{fig:completeness_curves}, except fake clumps are modeled as PSFs. The optimal smoothing length is 10 pixels (middle row), as was found when treating clumps as extended objects. }}
    \label{fig:completeness_curves_psf}
\end{figure*}

{To identify a complete sample of clumps under the assumption that they resemble PSFs, we perform the same routines to identify clumps and determine the completeness thereof as described in sections \ref{sec:clump_identification} and \ref{sec:completeness}, with only two exceptions as follows. First, aperture corrections are applied assuming the typical light profile of a clump is that of the PSF. Aperture corrections are 1.348, 1.307, 1.299, 1.336, and 1.523 in GOODS-N and 1.365, 1.328, 1.330, 1.356, and 1.485 in GOODS-S for the F090W, F115W, F150W, F200W, and F277W bandpasses, respectively. Second, to determine the completeness as a function of redshift and stellar mass, PSFs are inserted randomly into real images of galaxies, instead of S\'{e}rsic profiles convolved with PSFs. The resulting completeness curves are shown in Figure \ref{fig:completeness_curves_psf}. It is in the same format as Figure \ref{fig:completeness_curves}. The optimal smoothing length is 10 pixels, the same as that when treating clumps as extended objects. Complete samples of clumps can be detected at fainter minimum fractional luminosities (as low as $2\%$) when modeled as PSFs. This is largely because aperture corrections are smaller. }

\bibliography{sample631}{}
\bibliographystyle{aasjournal}



\end{document}